\documentclass[11pt,letterpaper,final]{iopart}

\usepackage{multicol,natbib}
\usepackage{amssymb}
\usepackage{graphicx,color,lscape,rotating}

\def\r2500{$r_{2500}$}

\def\apj{\it Astrophysical Journal\rm}
\def\mnras{\it Mon. Not. Royal Astron. Soc.\rm}

\def\sza{SZA}

\newcommand{\chandra}{\emph{Chandra}}
\newcommand{\wmap}{\emph{WMAP}}

\newcommand{\planck}{\emph{Planck}}
\def\apjl{\it Astrophysical Journal Letters\rm}
\def\apjs{\it Astrophysical Journal Supplements\rm}
\def\araa{\it Annual review of Astronomy and Astrophysics\rm}

\def\aap{\it Astronomy and Astrophysics\rm}

\def\altaffilmark#1{\textsuperscript{#1}}
\def\altaffiltext#1#2{%
\textsuperscript{#1}\hspace*{0.7mm}#2}%

\begin{document}

\title[Pressure Profiles in Relaxed Galaxy Clusters from X-ray and SZE Observations]{Comparison of Pressure Profiles of Massive Relaxed Galaxy Clusters using Sunyaev-Zel'dovich and X-ray Data}
\author{Massimiliano~Bonamente,\altaffilmark{1,2}
Nicole~Hasler,\altaffilmark{1}
Esra~Bulbul,\altaffilmark{1}
John~E.~Carlstrom,\altaffilmark{3,4,5}
Thomas~L.~Culverhouse,\altaffilmark{3}
Megan~Gralla,\altaffilmark{3}
Christopher~Greer,\altaffilmark{3}
David~Hawkins,\altaffilmark{6}
Ryan~Hennessy,\altaffilmark{3}
Marshall~Joy,\altaffilmark{2}
Jeffery~Kolodziejczak,\altaffilmark{2}
James~W.~Lamb,\altaffilmark{6}
David~Landry,\altaffilmark{1}
Erik~M.~Leitch,\altaffilmark{3}
Daniel~P.~Marrone,\altaffilmark{8}
Amber~Miller,\altaffilmark{9,10}
Tony~Mroczkowski,\altaffilmark{11}
Stephen~Muchovej,\altaffilmark{6}
Thomas~Plagge,\altaffilmark{3}
Clem~Pryke,\altaffilmark{3,4}
Matthew Sharp,\altaffilmark{3}
and David~Woody\altaffilmark{6}
}
\begin{footnotesize}
\altaffiltext{1}{Department of Physics, University of Alabama, Huntsville, AL 35899}
\altaffiltext{2}{Space Science-VP62, NASA Marshall Space Flight Center, Huntsville, AL 35812}
\altaffiltext{3}{Kavli Institute for Cosmological Physics and the Department of Astronomy and Astrophysics, University of Chicago, Chicago, IL 60637}
\altaffiltext{4}{Enrico Fermi Institute, University of Chicago, Chicago, IL 60637}
\altaffiltext{5}{Department of Physics, University of Chicago, Chicago, IL 60637}
\altaffiltext{6}{Owens Valley Radio Observatory, California Institute of Technology, Big Pine, CA 93513}
\altaffiltext{7}{NASA Goddard Space Flight Center, Greenbelt, MD 20771}
\altaffiltext{8}{Steward Observatory, University of Arizona, 933 North Cherry Avenue, Tucson, AZ 85721}
\altaffiltext{9}{Columbia Astrophysics Laboratory, Columbia University, New York, NY 10027}
\altaffiltext{10}{Department of Physics, Columbia University, New York, NY 10027}
\altaffiltext{11}{Department of Physics and Astronomy, University of Pennsylvania, Philadelphia, PA 19104}
\end{footnotesize}
\begin{abstract}
We present Sunyaev-Zel'dovich (SZ) effect
observations of a sample of 25 massive relaxed galaxy clusters
observed with the Sunyaev-Zel'dovich Array (SZA), an 8-element interferometer 
that is part of the Combined Array for Research in Millimeter-wave Astronomy (CARMA).
We perform an analysis of new SZA data and archival \chandra\ 
observations of this sample to investigate the integrated 
pressure---a proxy for cluster mass---determined 
from X-ray and SZ observations, two independent probes of the intra-cluster medium.
This analysis makes use of a model for the intra-cluster medium
introduced by Bulbul (2010) which can be applied simultaneously 
to SZ and X-ray data.  With this model, we estimate 
the pressure profile for each cluster using a joint analysis
of the SZ and X-ray data, and using the SZ data alone.  
We find that the integrated pressures 
measured from X-ray and SZ data are consistent.
This conclusion is in agreement with recent results 
obtained using \textit{WMAP} and \textit{Planck} data, 
confirming that SZ and X-ray observations of massive clusters detect 
the same amount of thermal pressure from the intra-cluster medium.
To test for possible biases introduced by our choice of model,
we also fit the SZ data using the universal pressure profile 
proposed by Arnaud (2010), and find consistency between the two
models out to $r_{500}$ in the pressure profiles and
integrated pressures. 
\end{abstract}

\maketitle
\section{Introduction}
The Sunyaev-Zel'dovich (SZ) effect \citep{sunyaev1972} is a spectral 
distortion of the cosmic microwave background (CMB) caused
by the scattering of CMB photons off the hot electrons of the intra-cluster medium (ICM).
Over the past two decades, SZ observations with both single-dish and interferometric instruments
have become routine \citep[e.g.,][]{birkinshaw1991,carlstrom1996,holzapfel1997,carlstrom2002},
and SZ surveys are now producing catalogs of newly-discovered clusters out to 
high redshift \citep[][]{vanderlinde2010,marriage2011,williamson2011,ade2011}.
SZ measurements are complementary to the X-ray measurements which have long been used to study
clusters,
but only in recent years have sufficiently large samples of objects been observed in the
SZ to permit a rigorous comparison between these two techniques
\citep[e.g.,][]{reese2002,bonamente2006,laroque2006}. 

The SZ effect causes a perturbation $\Delta T$ of the CMB temperature
$T_{CMB}$ given by
\begin{equation}
\frac{\Delta T}{T_{CMB}} = f(x) \int \! \sigma_T n_e \frac{k T_e}{m_e c^2} d\ell = f(x) y ,
\label{eq:compy}
\end{equation}
where $f(x)$ is the frequency dependence of the SZ effect \citep[e.g.,][]{laroque2006};
$\sigma_T$ is the Thomson cross-section; $n_e$, $T_e$ and $m_e$ are the number density,
temperature, and mass of the electrons, respectively; $k$ is the Boltzmann constant; 
$c$ the speed of light; and the integral is along the line of sight $\ell$.
At a given frequency, the amplitude of the effect depends linearly upon the 
Compton $y$ parameter, which is defined implicitly in Equation~\ref{eq:compy}.
Note that the $y$ parameter is proportional to the ICM pressure integrated along the
line of sight.  At frequencies below 218~GHz, the SZ effect causes  a decrement in the 
CMB temperature in the direction of the cluster.
The integral of $y$ over the solid angle $\Omega$ subtended by the cluster,
known as the (cylindrical) integrated Compton $y$ parameter
$Y_{\rm cyl}=\int \! y \, d \Omega$, is expected to be a good proxy 
for cluster total mass since it traces the thermal energy content 
of the cluster gas.  Alternatively, the Compton
$y$ parameter can be integrated spherically,
\begin{equation}
Y_{\rm sph}(r_{500}) = \frac{1}{D_A^2} \left( \frac{k \sigma_T}{m_e c^2} \right)\int  n_e T_e dV
\label{eq:Y}
\end{equation}
where the volume $V$ is a sphere centered on the cluster and $D_A$ is the angular
diameter distance. 

X-ray data can also be used to constrain the density and temperature---and thus
the pressure---of the ICM.  Over the past decade, several groups have investigated
the consistency between X-ray and SZ pressure measurements.
Early measurements of the SZ signal from \wmap\ by, e.g., \cite{lieu2006}, \cite{bielby2007},
detected an SZ signal at a lower level than expected.
\cite{atrio-barandela2008} showed that the isothermal beta model
leads to an electron pressure profile that exceeds the measured values at large radii
by a factor of few, and that the baryon profile is consistent with a model based on the
\cite{navarro1997} matter profile. \cite{diego2010} also showed that
 contamination by compact radio sources may have led to underestimates     
of the SZ effect flux decrements in the \wmap\ data.
More recent comparisons of \chandra\ X-ray data to stacked data from \wmap\
and \planck\ \citep{melin2011,Planck2011-sz} have found
consistency between SZ and X-ray measurements for large samples of clusters.
\cite{komatsu2011} also analyzed a sample of massive nearby clusters 
individually resolved by \wmap, again finding good agreement with
X-ray predictions.

In this paper, we present Sunyaev-Zel'dovich Array (\sza) observations 
of the \cite{allen2008} sample of massive relaxed galaxy clusters, together 
with archival \chandra\ X-ray observations that are
available for all clusters in this sample.
The sensitivity and resolution of our data permit us to measure the pressure profile
and the integrated pressure out to $r_{500}$---the radius within which the average
cluster density is 500 times the critical density---for each 
cluster individually, without the need to resort to scaling relations between
the X-ray luminosity and mass \citep[as was done by][for example]{melin2011,Planck2011-sz}.
We use the \citet{bulbul2010} model of the cluster pressure, density, and temperature. 
Since this model has a consistent parameterization for all thermodynamic quantities, it is 
especially well-suited for joint X-ray and SZ analysis.  As a cross-check against
model-dependent biases, we also fit the SZ data using the model of \citet{arnaud2010}
based on the numerical simulations of \cite{nagai2007b}.  We find consistency to
within our measurement uncertainties both between the X-ray and SZ measurements,
and between the \citet{bulbul2010} and \citet{arnaud2010} models.

The paper is structured as follows: Section~\ref{sec:sample} describes
our observations and our sample, 
Section~\ref{sec:analysis} presents our joint analysis technique, 
Section~\ref{sec:Y} describes our method of measuring
the integrated $Y_{\rm sph}(r_{500})$ parameter (defined in Equation~\ref{eq:Y}), 
Section~\ref{sec:discussion} presents and discusses our results, 
and our conclusions are presented in Section~\ref{sec:conclusions}.

\section{Observations}
\label{sec:sample}

The \sza\ is an eight-element interferometer designed 
to detect and image the SZ effect from clusters at z$>0.1$,
and is part of the Combined Array for Research in 
Millimeter-wave Astronomy (CARMA). 
The array is equipped with 30 and 90~GHz receivers; all \sza\
observations presented in this paper were taken at 30~GHz. 
At this frequency, the 3.5~m diameter \sza\ telescopes 
have a field-of-view (or primary beam) of 10.$^\prime$7 FWHM.
Interferometric data are proportional to the Fourier transform of the sky brightness. 
These \emph{visibility} data, denoted $V(u,v)$, are sampled at Fourier-plane coordinates 
$(u,v)$ corresponding to the projected separation of pairs of telescopes 
(or \emph{baselines}), as viewed by the source at the time of observation. 
At the time of the observations discussed in this work, the \sza\ antennas were 
arranged in a hybrid configuration, with six closely 
spaced telescopes and two ``outriggers'' located $\sim$50~m from the inner array. 
The inner six telescopes probe small $(u,v)$ Fourier modes, 
sampling the angular scales where the SZ signal is largest 
for moderate- to high-redshift clusters ($1-6^\prime$).
Baselines involving the outriggers are sensitive to angular scales down to $\sim20^{\prime\prime}$ 
and are used to constrain the positions and fluxes of unresolved radio sources.

Of the 42 clusters in the \citet{allen2008} sample of massive relaxed
galaxy clusters, the \sza\ has observed the 31 objects
above $\delta > -15^{\circ}$ at redshift $z\geq0.09$.  The declination 
restriction is imposed by the latitude of the observatory in the 
Owens Valley, California ($37^\circ14'02''$N, $118^\circ16'56''$W), 
while the exclusion of clusters at low redshift is due to the inability
of an interferometer to constrain scales larger than that which 
the shortest antenna spacing can probe at the lowest frequency 
band.  The largest angular wavelength measured by the \sza\ is 
$10.9^\prime$, which for massive low-redshift clusters is 
generally smaller than $2 r_{500}/D_A$.
Of these 31 clusters observed with the \sza,
Abell~2390 and Abell~611 were excluded from this analysis
because they did not have available local background in their 
\chandra\ ACIS-S X-ray observations.
Three additional clusters---3C295, ClJ1415.2+3612, and Abell~963---were 
discarded because of extended or otherwise difficult-to-remove
radio source contamination, and one---RXJ0439.0+0521---because 
of a pointing error.  

Our sample therefore consists of 25 clusters. 
The synthesized beam of the long (short) baseline data 
for this sample is approximately 15-30$^{\prime\prime}$ 
(90-180$^{\prime\prime}$), and the average $rms$ noise 
in the maps is $\sim0.25-0.30$ mJy.
In all cases, the \chandra\ data provide 
spatially resolved X-ray spectroscopy and sub-arcsecond imaging. 
A summary of the data is provided in Table \ref{tab:infoX}.
\begin{table}
\scriptsize
\centering
\caption{Sample of massive and relaxed clusters from the \cite{allen2008} sample with
high resolution Sunyaev-Zel'dovich effect \sza\ observations.}
\begin{tabular}{lcccc|cccc}
\hline
\hline
Cluster & z &  {R.A.} & {Dec.} & $N_H$ & \sza  &  \multicolumn{3}{c}{\chandra}  \\
                & & (J2000) & (J2000) & ($10^{20}$ cm$^{-2}$)$^a$  & 
(hrs)$^b$ &  ACIS& ObsID  & (ks)$^c$ \\
\hline
MACSJ0159.8-0849     & 0.40 & 01 59 49.5  & -08 50 02 & 2.06 & 21.2 & I & 3265 & 16.4 \\ 
		     & 	    & 		  & 	      &      &      & I & 6106 & 34.3 \\
		     & 	    & 		  & 	      &      &      & I & 9376 & 19.5 \\
Abell~383 	     & 0.19 & 02 48 03.4  & -03 31 44 & 3.40 & 25.0 & I & 524  & 9.9  \\
		     &      &             &           &      &      & I & 2320 & 18.5 \\
MACSJ0329.7-0212     & 0.45 & 03 29 41.7  & -02 11 48 & 3.43 & 8.1  & I & 6108 & 32.7 \\
		     &      & 		  & 	      &      &      & I & 3257 & 9.6  \\
		     & 	    & 		  & 	      &      &      & I & 3582 & 19.3 \\
Abell~478 	     & 0.09 & 04 13 25.2  & +10 27 52 & 34.29& 37.1 & I & 6102 & 10.0 \\
MACSJ0429.6-0253     & 0.40 & 04 29 36.1  & -02 53 08 & 3.23 & 22.1 & I & 3271 & 23.2 \\
3C186 	             & 1.06 & 07 44 17.5  & +37 53 17 & 5.11 & 13.7 & S & 9407 & 66.3 \\
		     & 	    & 		  & 	      &      &      & S & 9408 & 39.6 \\
		     & 	    & 		  & 	      &      &      & S & 9774 & 75.1 \\
		     &      &             &           &      &      & S & 9775 & 15.9 \\
MACSJ0744.9+3927     & 0.69 & 07 44 52.9  & +39 27 26 & 5.66 & 12.7 & I & 6111 & 49.5 \\
		     & 	    &             &           &      &      & I & 3197 & 20.2 \\
		     & 	    & 		  & 	      &      &      & I & 3585 & 19.7 \\
MACSJ0947.2+7623     & 0.34 & 09 47 13.1  & +76 23 14 & 2.28 & 11.5 & I & 2202 & 11.7 \\
Zwicky~3146          & 0.29 & 10 23 39.6  & +04 11 10 & 2.46 & 6.8  & I & 909  & 45.2 \\
		     &      & 		  & 	      &      &      & I & 9371 & 36.3 \\
MACSJ1115.8+0129     & 0.35 & 11 15 52.0  & +01 29 58 & 4.34 & 26.2 & I & 9375 & 39.6 \\
MS1137.5+6625        & 0.78 & 11 40 22.2  & +66 08 14 & 0.95 & 19.6 & I & 536  & 109.6\\
Abell~1413           & 0.14 & 11 55 18.2  & +23 24 19 & 3.60 & 43.2 & I & 5003 & 66.6 \\
		     & 	    & 		  & 	      &      &      & I & 1661 & 9.1  \\
		     & 	    & 		  & 	      &      &      & I & 5002 & 34.4 \\
ClJ1226.9+3332       & 0.89 & 12 26 58.2  & +33 32 47 & 1.83 & 16.0 & I & 5014 & 31.6 \\
		     &      &		  & 	      &      &      & I & 3180 & 29.9 \\
MACSJ1311.0-0311     & 0.49 & 13 11 01.7  & -03 10 38 & 1.82 & 4.6  & I & 6110 & 63.0 \\
		     &      & 		  & 	      &      &      & I & 3258 & 13.1 \\
		     & 	    & 		  & 	      &      &      & I & 9381 & 29.0 \\
RXJ1347.5-1145       & 0.45 & 13 47 30.6  & -11 45 10 & 4.60 & 25.7 & I & 3592 & 54.8 \\
Abell~1835           & 0.25 & 14 01 02.0  & +02 52 40 & 2.04 & 9.0  & I & 6880 & 117.9\\
		     &      &             &           &      &      & I & 6881 & 36.8 \\
		     & 	    & 		  & 	      &      &      & I & 7370 & 40.0 \\
MACSJ1423.8+2404     & 0.54 & 14 23 47.9  & +24 04 42 & 2.20 & 6.5  & I & 1657 & 18.2 \\
MACSJ1427.3+4408     & 0.49 & 14 27 16.3  & +44 07 29 & 1.19 & 17.4 & I & 6112 &  8.8 \\
		     &      &             &           &      &      & I & 9380 & 25.8 \\
		     &      &             &           &      &      & I & 9808 & 14.9 \\
RXJ1504.1-0248       & 0.21 & 15 04 07.5  & -02 48 16 & 5.97 & 9.2  & I & 5793 & 39.2 \\
		     &      &             &           &      &      & I  & 4935 & 11.9 \\
MACSJ1532.9+3021     & 0.36 & 15 32 53.8  & +30 20 58 & 2.30 & 14.6 & I & 1665 & 8.2  \\
MACSJ1621.6+3810     & 0.46 & 16 21 24.9  & +38 10 08 & 1.13 & 44.0 & I & 6172 & 29.2 \\
		     &      &             &           &      &      & I & 3254 & 9.6  \\
		     &      &             &           &      &      & I & 6109 & 36.7 \\
		     &      &             &           &      &      & I & 9379 & 29.7 \\
		     &      &             &           &      &      & I & 10785 & 29.7 \\
Abell~2204           & 0.15 & 16 32 46.9  & +05 34 31 & 5.67 & 19.6 & I & 7940 & 76.9 \\
MACSJ1720.3+3536     & 0.39 & 17 20 16.8  & +35 36 25 & 3.46 & 36.2 & I & 6107 & 29.1 \\
		     &      &             &           &      &      & I & 3280 & 20.6 \\
		     &      &             &           &      &      & I & 7718 &  7.0 \\
RXJ2129.6+0005       & 0.23 & 21 29 40.0  & +00 05 18 & 3.63 & 24.5 & I & 552  & 10.0 \\
Abell~2537           & 0.29 & 23 08 22.2  & -02 11 28 & 4.62 & 24.8 & I & 9372 & 38.5 \\
\hline
\hline
\label{tab:infoX}
\end{tabular}

$a$: {$N_H$ is HI Galactic column density.}\\
$b$: {\sza\ exposure is  unflagged, on-source time}\\
$c$: {\chandra\ exposure is unflagged, on-source time}
\end{table}

Radio sources detected in the cluster fields are reported in Table~\ref{tab:radio-ps}.
For each cluster field, we use the NRAO VLA Sky Survey (NVSS)
 and  Faint Images of the Radio Sky at Twenty-centimeters 
(FIRST) 1.4~GHz catalogs as a reference for locating compact radio sources within 10$'$ of the cluster center.
Most radio sources in our observations have counterparts in the
FIRST survey, which has an $rms$ noise of 0.15 mJy at 1.4 GHz. Inverted spectrum sources
that may be present at 30~GHz may not have counterparts at 1.4 GHz,
but fortunately they comprise a small fraction of the source population \citep{muchovej2010}.

For all 25 clusters in our sample we have available archival \chandra\ X-ray 
observations \citep{allen2008}.
Event files for all cluster observations and additional blank-sky composite event files
used for background subtraction were reduced
using CIAO 4.3.1 and CALDB 4.3.
X-ray spectra are extracted in several annular regions for each cluster,
centered at the peak of the X-ray emission.
Emphasis is placed on the removal of periods of high background, and on the
modeling of soft X-ray residuals that may be present after the
subtraction of the blank-sky background. The method of analysis
of the \chandra\ data and examples of the temperature and surface brightness
profiles can be found in \cite{bulbul2010} and \cite{hasler2011}.
More details on the \chandra\ data for all clusters in this sample 
will be shown in a forthcoming paper,
in which we will present the  measurement of
the gas mass fraction from the X-ray observations (Hasler et al. in prep.).

\begin{figure}
\centering
\includegraphics[trim= 0.5in 1.7in 0.5in 2.2in, angle=-90,width=0.28\textwidth]{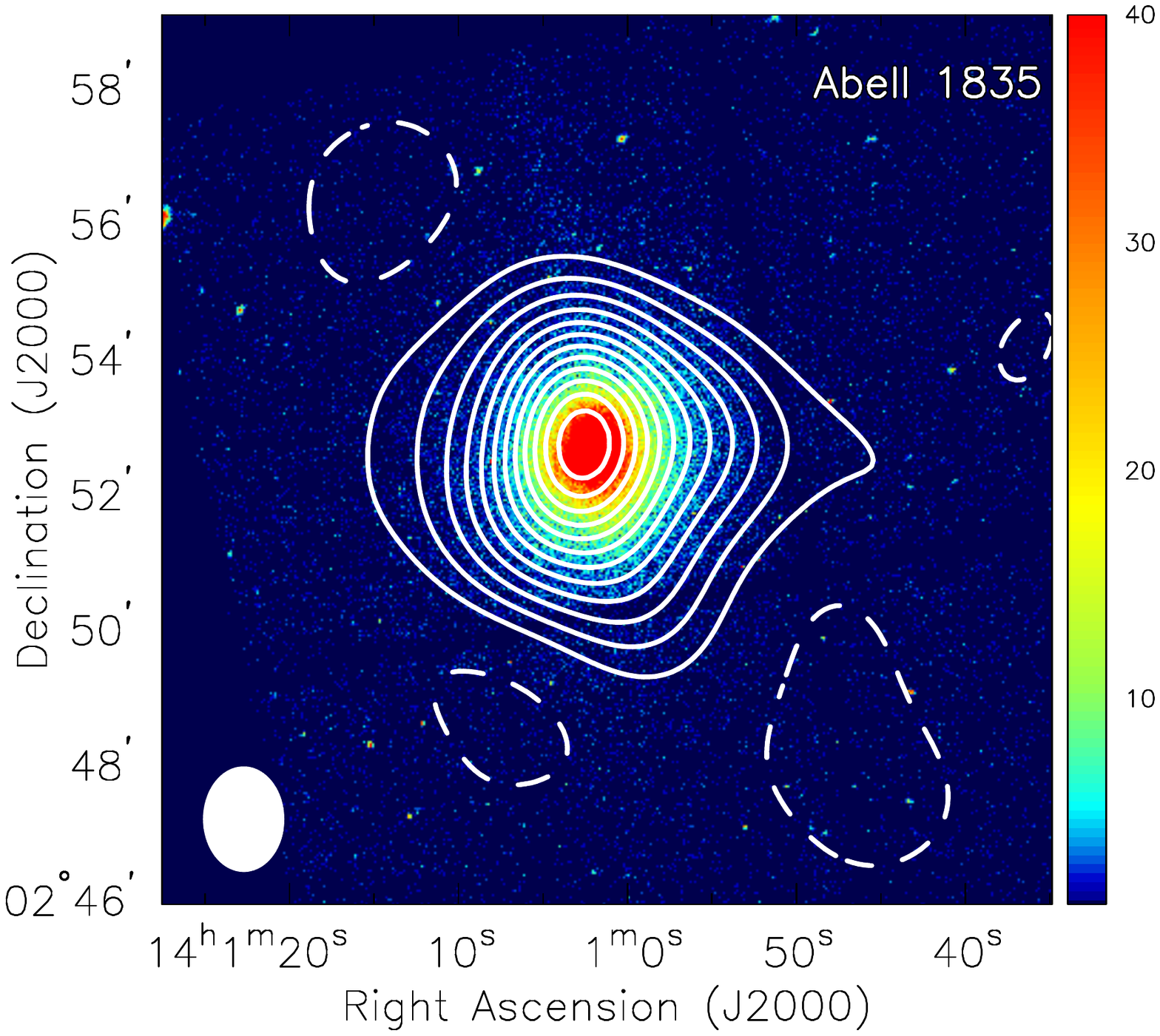}
\includegraphics[trim= 0.5in 1.7in 0.5in 2.2in, angle=-90,width=0.28\textwidth]{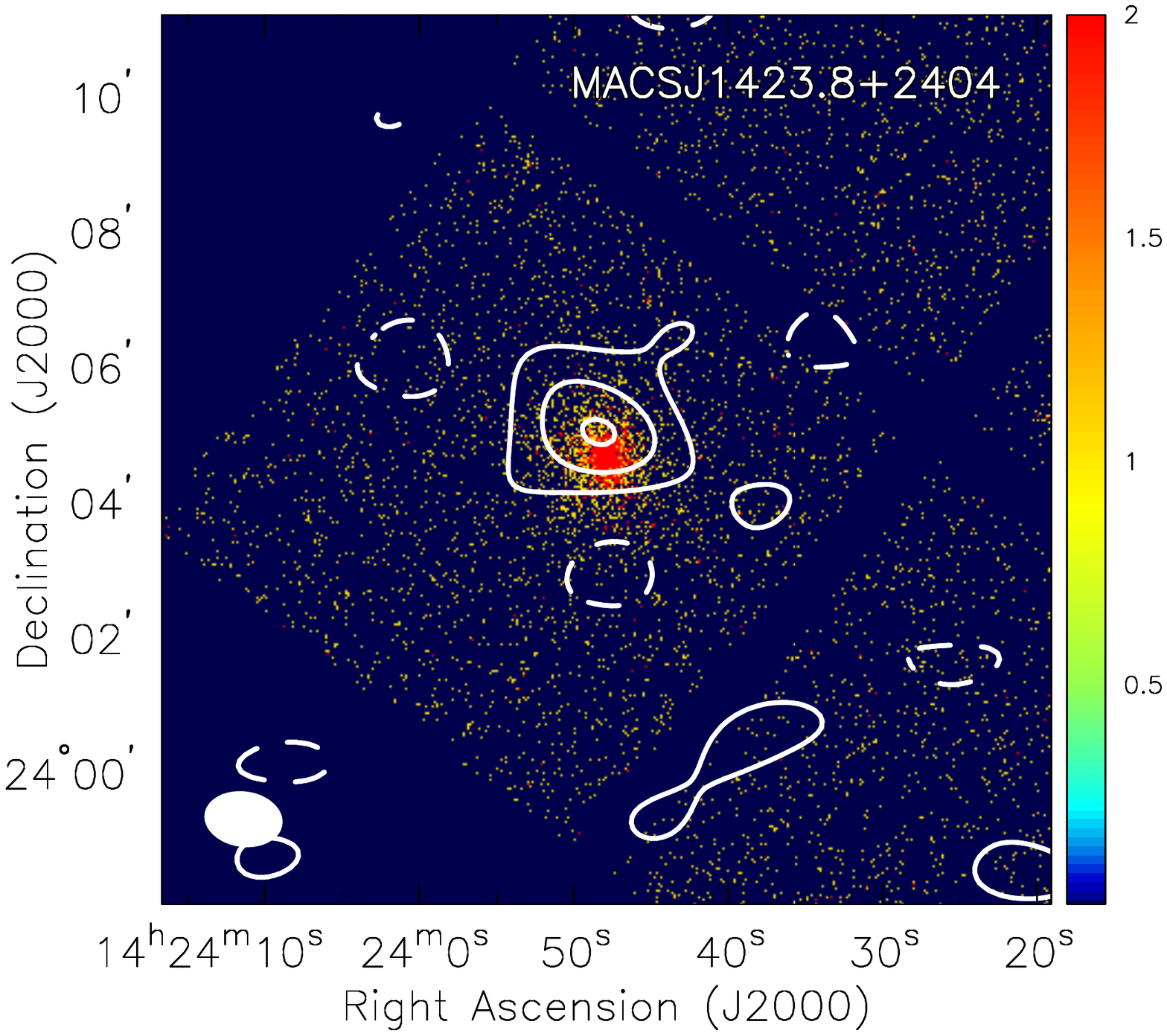}
\includegraphics[trim= 0.5in 1.7in 0.5in 2.2in, angle=-90,width=0.28\textwidth]{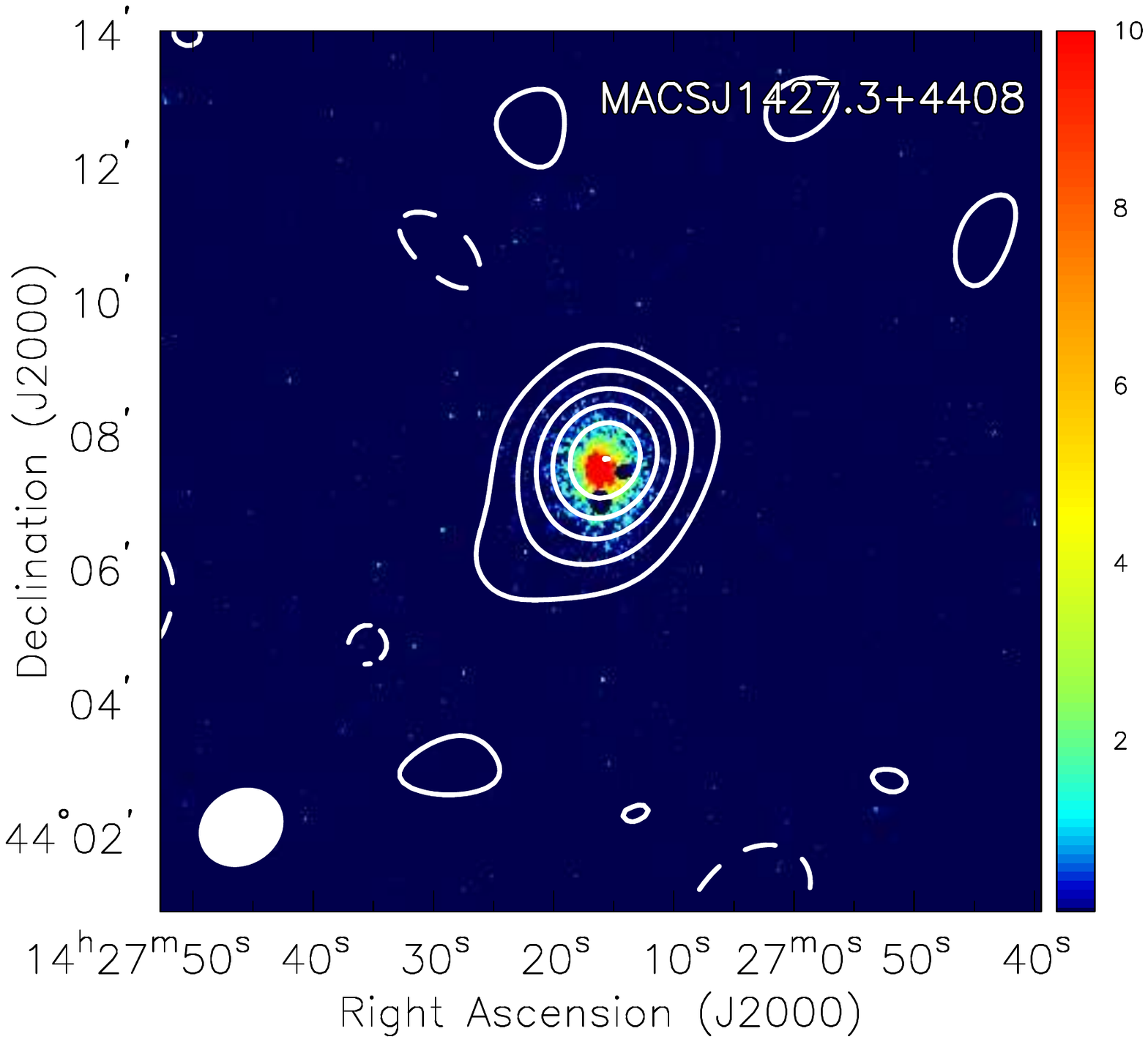}
\includegraphics[trim= 0.5in 1.7in 0.5in 2.2in, angle=-90,width=0.28\textwidth]{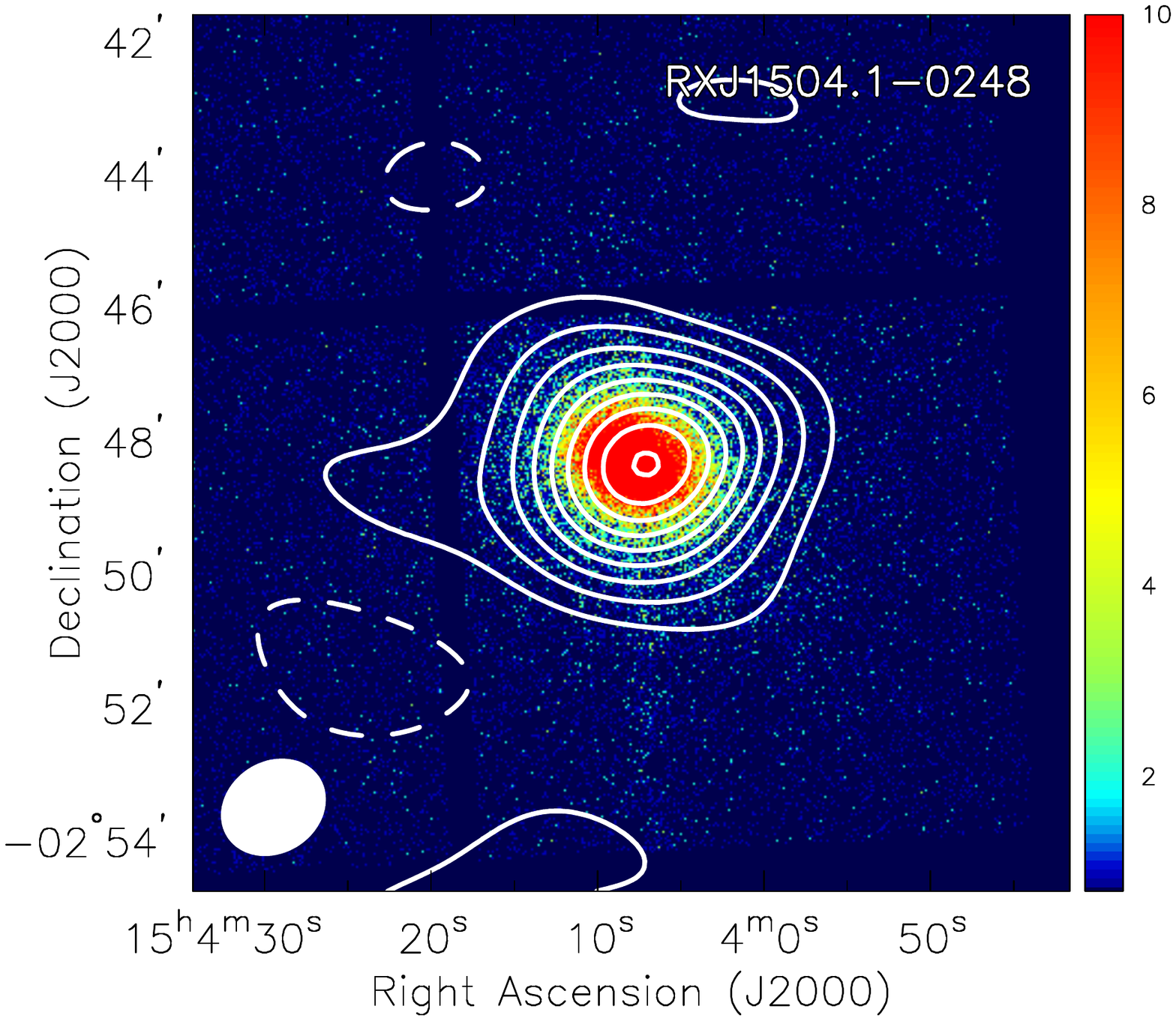}
\includegraphics[trim= 0.5in 1.7in 0.5in 2.2in, angle=-90,width=0.28\textwidth]{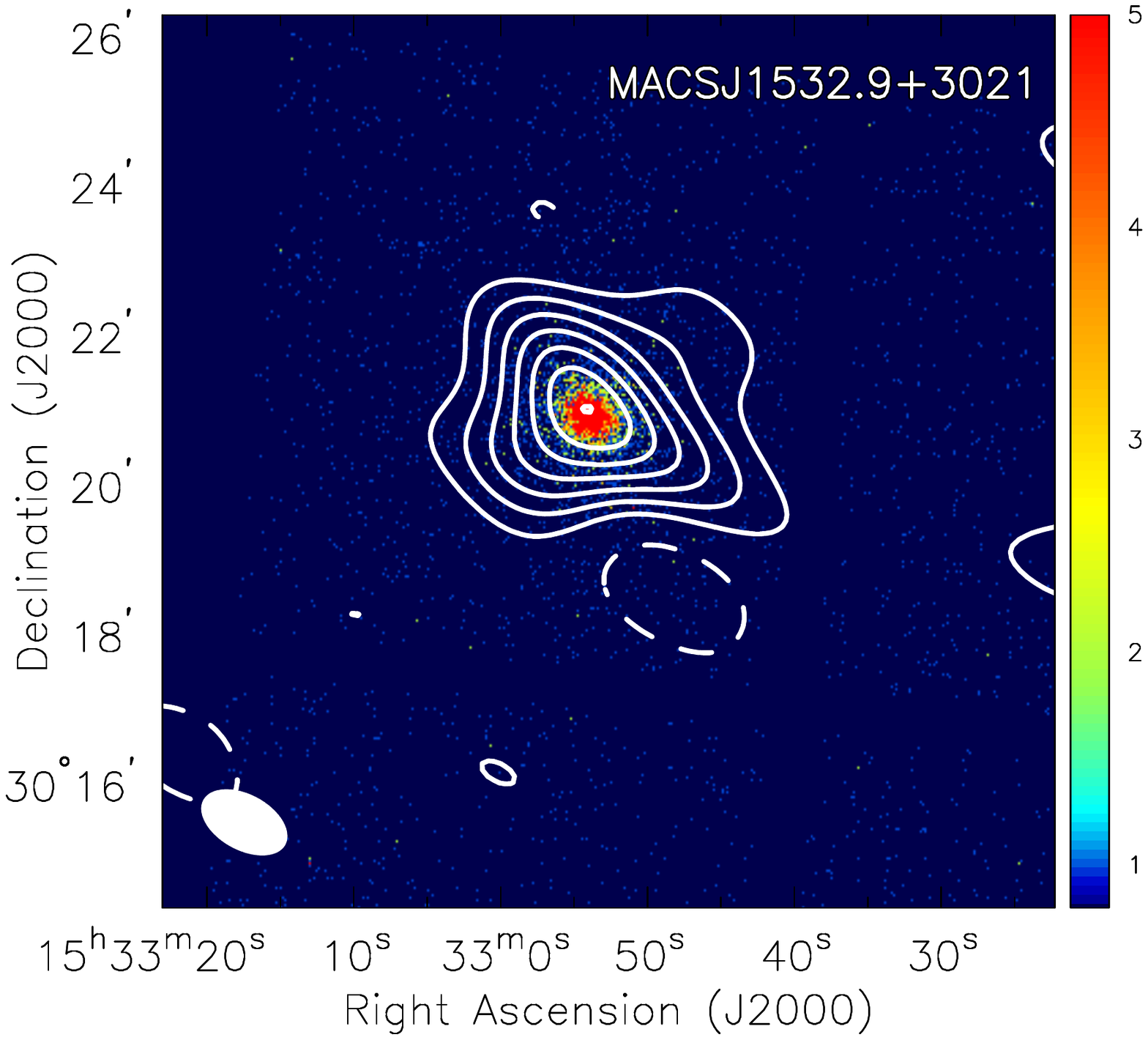}
\includegraphics[trim= 0.5in 1.7in 0.5in 2.2in, angle=-90,width=0.28\textwidth]{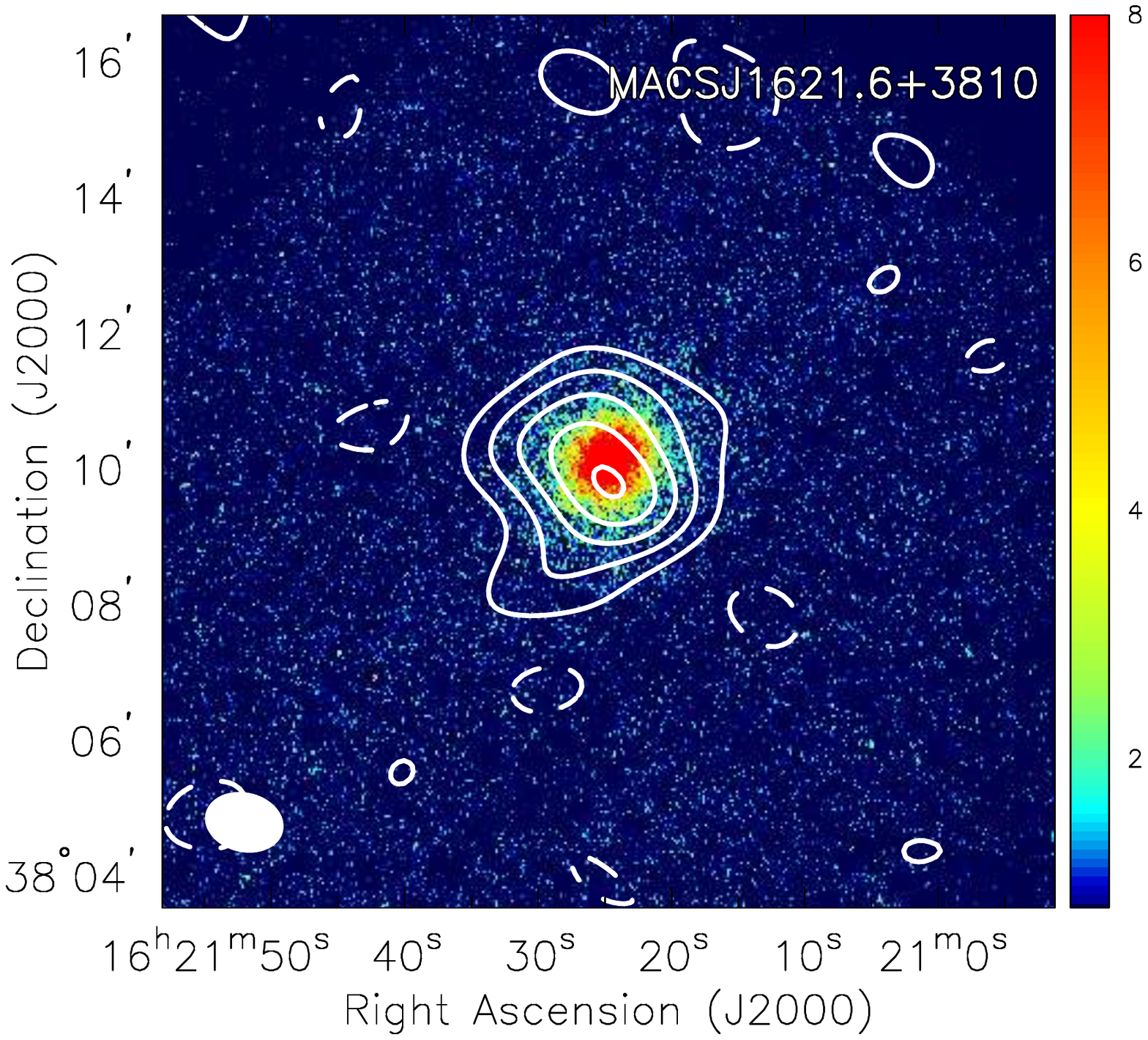}
\includegraphics[trim= 0.5in 1.7in 0.5in 2.2in, angle=-90,width=0.28\textwidth]{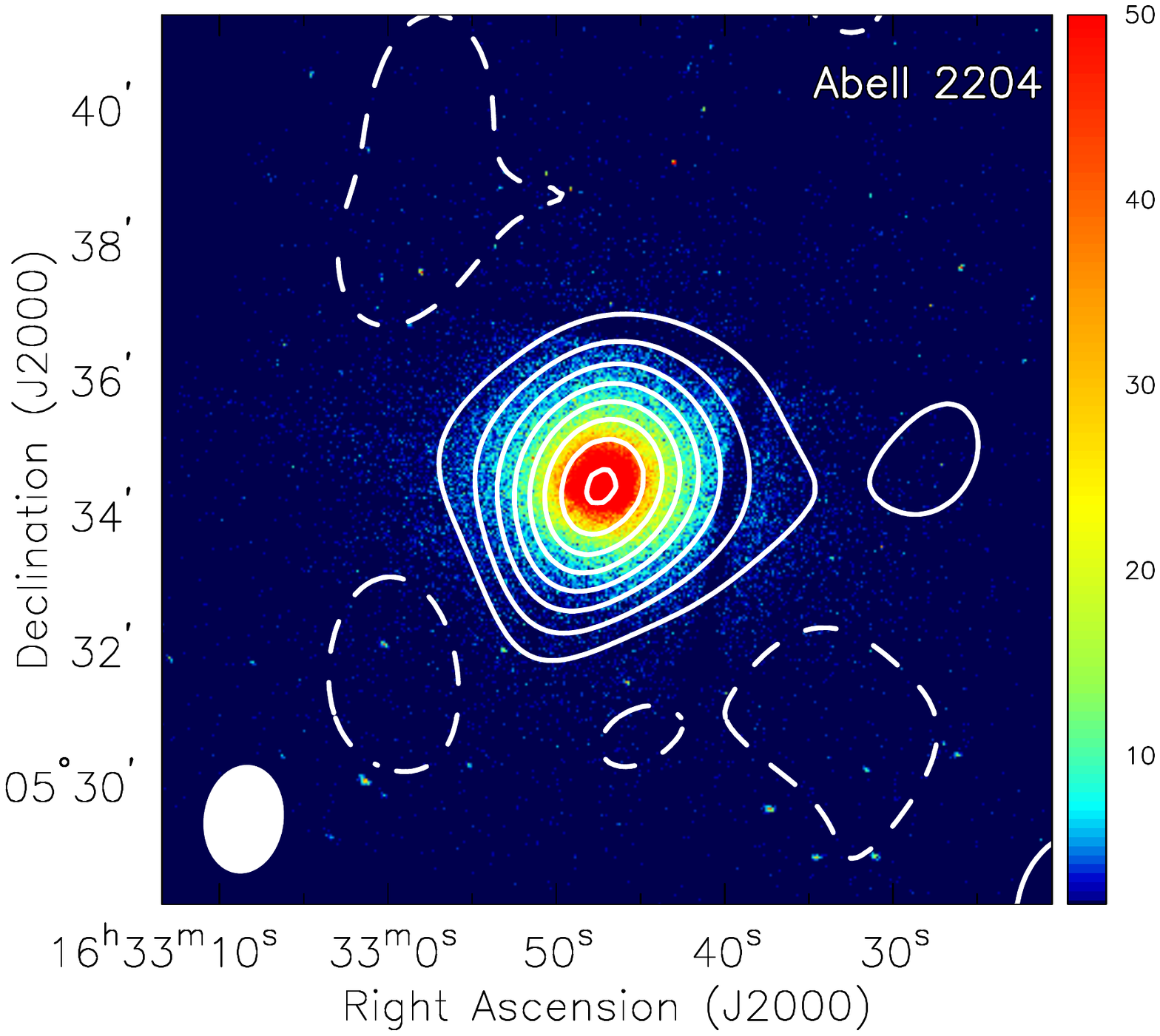}
\includegraphics[trim= 0.5in 1.7in 0.5in 2.2in, angle=-90,width=0.28\textwidth]{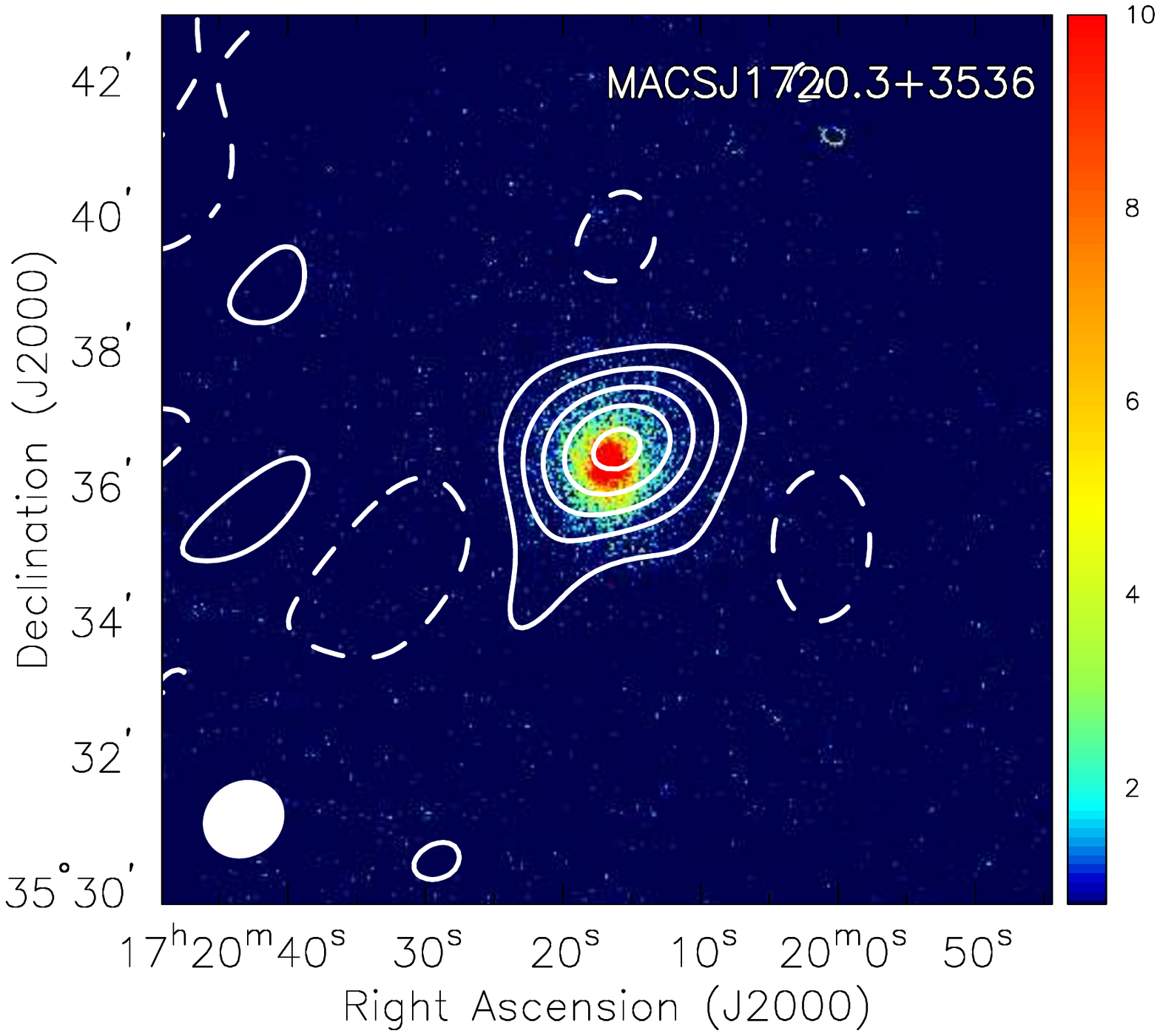}
\includegraphics[trim= 0.5in 1.7in 0.5in 2.2in, angle=-90,width=0.28\textwidth]{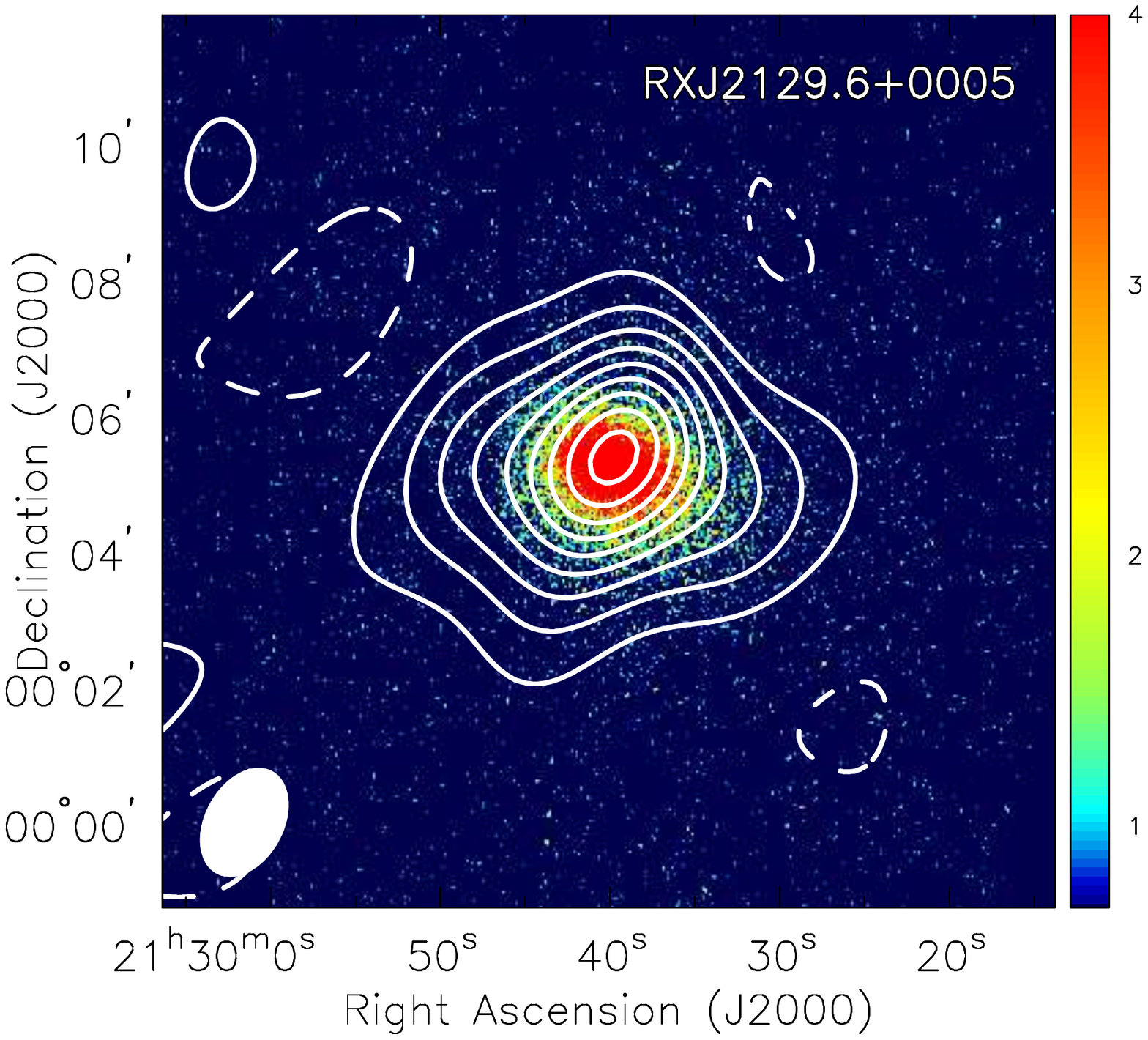}
\includegraphics[trim= 0.5in 1.7in 0.5in 2.2in, angle=-90,width=0.28\textwidth]{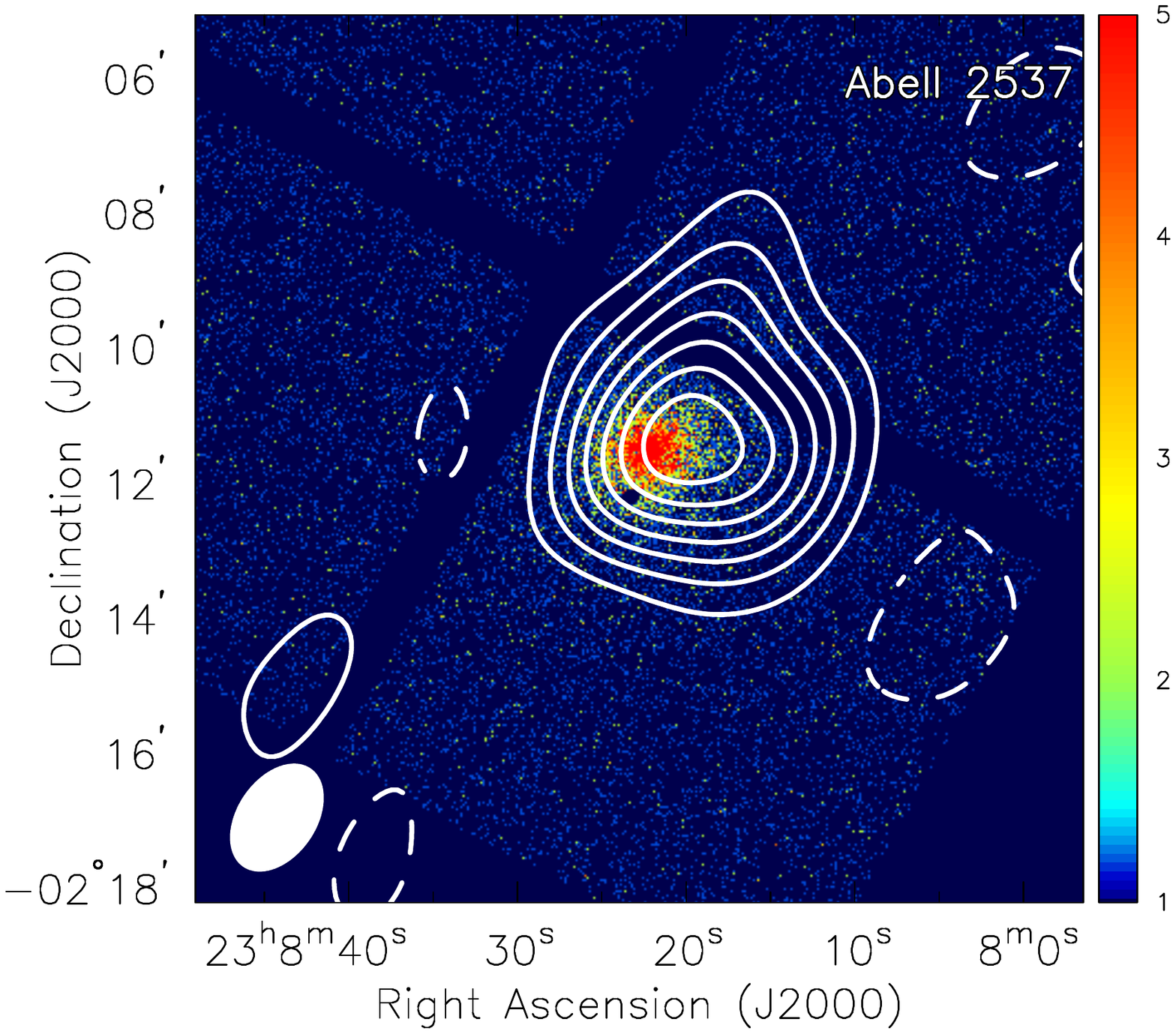}
\caption{
\chandra\ images in the 0.7-7~keV energy range. The color bars reflect the number of counts detected by \chandra.
SZ contour levels are
(+2,-2,-4,-6,-8,...) times the rms noise in the short baseline data, after
removal of radio sources; solid contours are for negative levels, and dashed contours are positive levels.
The elliptical Gaussian approximation to the synthesised beam of the SZ
observations are shown in the lower left corner.
 \label{fig:images}}
\end{figure}

In Figure~\ref{fig:images}, we show the raw \chandra\ X-ray images (binned in the 
0.7--7~keV energy band) for each of the 25 clusters, with contours obtained from
the short baseline point source-removed \sza\ data overlaid.

\begin{table}[!h]
\centering
\scriptsize
\caption{SZ Centroids and Radio Source Locations for the SZA Observations}
\begin{tabular}{lccccrrrrrr}
 & & \multicolumn{2}{c}{\underline{SZ Centroid}} & \multicolumn{4}{c}{\underline{30 GHz Source}} & \multicolumn{2}{c}{\underline{1.4 GHz Flux (mJy)}}\\
Cluster & z & $\alpha$(J2000) & $\delta$(J2000) & src & $\Delta \alpha$ ($^{\prime\prime}$)$^a$ 
& $\Delta \delta$ ($^{\prime\prime}$)$^a$ &  Flux (mJy) & NVSS & FIRST \\
\hline
 MACSJ0159.8-0849 & 0.40 & 01:59:51.5 & -08:50:06.9 & 1 & -31.3  & 8.0    &  84.4${\pm0.2}$ & 36.7  & 31.4 \\
 Abell~383        & 0.19 & 02:48:03.5 & -03:31:55.8 & 1 & -1.5   & 11.5   &   4.3${\pm0.2}$ & 40.9  & - \\
                  & ~    &            &             & 2 & 276.7  & -149.2 &   7.5${\pm0.3}$ & 54.9  & - \\
 MACSJ0329.7-0212 & 0.45 & 03:29:40.3 & -02:11:44.5 & 1 & 263.2  & -97.2  &  12.7${\pm0.4}$ & 37.2  & - \\
 Abell~478        & 0.09 & 04:13:25.0 & +10:27:50.8 & 1 & 195.4  & 18.1   &   2.9${\pm0.1}$ & 47.7  & - \\
                  & ~    &            &             & 2 & 3.3    & 4.1    &   2.3${\pm0.1}$ & 36.9  & - \\
 MACSJ0429.6-0253 & 0.40 & 04:29:35.6 & -02:53:01.6 & 1 & 7.0    & -7.7   &  18.2${\pm0.2}$ & 138.8 & - \\
                  & ~    &            &             & 2 & -285.4 & 53.9   &   3.2${\pm0.2}$ & 18.3  & - \\
 3C186            & 1.06 & 07:44:14.8 & +37:53:21.2 & 1 & 40.8   & -3.3   &  22.6${\pm0.2}$ & 1236.4 & 1244.9 \\
                  & ~    &            &             & 2 & 81.6   & -94.3  &  10.6${\pm0.2}$ & 105.4 & 49.2 \\
 MACSJ0744.9+3927 & 0.69 & 07:44:52.2 & +39:27:34.6 & 1 & -215.6 & 286.7  &   2.5${\pm0.5}$ & -     & 4.4 \\
 MACSJ0947.2+7623 & 0.34 & 09:47:12.4 & +76:23:03.0 & 1 & 11.1   & 11.2   &   2.5${\pm0.3}$ & 21.7  & -\\
 Zwicky~3146      & 0.29 & 10:23:38.9 & +04:11:27.7 & 1 & 92.4   & -48.5  &   5.0${\pm0.3}$ & 95.8  & 56.7 \\
                  & ~    &            &             & 2 & 8.9    & -17.6  &   1.7${\pm0.2}$ & 7.1   & 2.0 \\
                  & ~    &            &             & 3 & -46.2  & -138.7 &   2.2${\pm0.2}$ & 31.5  & 15.1 \\
 MACSJ1115.8+0129 & 0.35 & 11:15:52.2 & +01:29:50.6 & 1 & 128.7  & -360.5 &   2.7${\pm0.2}$ & 11.5  & 10.5 \\
                  & ~    &            &             & 2 & 162.2  & -95.5  &   3.7${\pm0.2}$ & -     & 2.1 \\
                  & ~    &            &             & 3 & 139.7  & -64.8  &   1.8${\pm0.3}$ & -     & - \\
                  & ~    &            &             & 4 & -3.1   & 3.0    &   1.4${\pm0.4}$ & 6.2   & 5.6 \\
                  & ~    &            &             & 5 & -168.9 & -71.2  &   2.7${\pm0.2}$ & 10.3  & 6.4 \\
 MS1137.5+6625    & 0.78 & 11:40:22.8 & +66:08:13.2 & - & -      & -      &   -             & -     & - \\
 Abell~1413       & 0.14 & 11:55:17.5 & +23:24:04.0 & 1 & -117.0 &  135.5 &   2.1${\pm0.1}$ & 28.1  & 19.8\\
                  & ~    &            &             & 2 & -386.0 & -185.2 &   2.8${\pm0.3}$ & -     & - \\
 CLJ1226.9+3332   & 0.89 & 12:26:57.7 & +33:32:51.8 & 1 & 263.4  & -46.0  &   3.9${\pm0.2}$ & 29.8  & 23.2 \\
 MACSJ1311.0-0311 & 0.49 & 13:11:02.2 & -03:10:47.0 & - & -      & -      &   -             & -     & - \\
 RXJ1347.5-1145   & 0.45 & 13:47:31.4 & -11:45:16.1 & 1 & -11.4  & 6.3    &   8.7${\pm0.2}$ & 45.9  & - \\
                  & ~    &            &             & 2 & -339.1 & 251.5  &   6.9${\pm0.4}$ & 365.8 & - \\
                  & ~    &            &             & 3 & -53.3  & 279.5  &   2.6${\pm0.3}$ & 5.1   & - \\
 Abell~1835       & 0.25 & 14:01:02.2 & +02:52:34.4 & 1 & -1.5   & 9.2    &   2.9${\pm0.3}$ & 39.3  & 31.3\\
                  & ~    &            &             & 2 & -29.0  & -47.7  &   1.0${\pm0.3}$ & -     & - \\
 MACSJ1423.8+2404 & 0.54 & 14:23:48.6 & +24:05:13.6 & 1 & -11.8  & -31.5  &   2.0${\pm0.2}$ & 8.0   & 5.2 \\
 MACSJ1427.3+4408 & 0.49 & 14:27:15.8 & +44:07:41.4 & 1 & 4.8    & -10.8  &  16.4${\pm0.2}$ & 47.9  & 41.3 \\
                  & ~    &            &             & 2 & 33.6   & 206.8  &   1.1${\pm0.2}$ & 8.6   & 8.2 \\
 RXJ1504.1-0248   & 0.21 & 15:04:07.1 & -02:48:17.8 & 1 & 5.7    &  1.3   &  15.9${\pm0.2}$ & 60.5  & 40.8 \\
 MACSJ1532.9+3021 & 0.36 & 15:32:54.0 & +30:20:59.0 & 1 & -39.3  & -72.9  &   5.7${\pm0.2}$ & 7.9   & 6.0 \\
                  & ~    &            &             & 2 &  -2.7  & 0.3    &  3.2${\pm0.2}$ & 22.8  & 15.2 \\
                  & ~    &            &             & 3 & -82.8  & -128.4 &   1.3${\pm0.2}$ & 18.0  & 4.1 \\
 MACSJ1621.6+3810 & 0.46 & 16:21:25.3 & +38:09:56.9 & - & -      & -      &   -             & -     & - \\
 Abell~2204       & 0.15 & 16:32:47.2 & +05:34:34.7 & 1 & -3.6   & -1.5   &   7.0${\pm0.2}$ & 69.3  & 57.9 \\
                  & ~    &            &             & 2 & -421.8 & -362.8 &  21.6${\pm0.2}$ & 41.6  & - \\
                  & ~    &            &             & 3 & 191.0  & -132.8 &   0.7${\pm0.1}$ & 12.2  & 1.2 \\
 MACSJ1720.3+3536 & 0.39 & 17:20:16.2 & +35:36:36.0 & 1 & 650.3  &  340.2 & 167.7${\pm0.2}$ & -     & - \\
                  & ~    &            &             & 2 & 10.3   & -9.4   &   1.8${\pm0.4}$ & 18.0  & 16.8 \\
 RXJ2129.6+0005   & 0.23 & 21:29:40.2 & +00:05:20.9 & 1 & -3.2   &  0.4   &   2.6${\pm0.2}$ & 25.4  & 23.8\\
                  & ~    &            &             & 2 & 228.1  & 160.9  &   3.1${\pm0.2}$ & 34.3 & 6.6 \\
 Abell~2537       & 0.29 & 23:08:19.2 & -02:11:19.0 & 1 & 138.6  & 437.2  &   8.4${\pm0.9}$ & 69.9 & 58.6\\
\hline
\hline
\label{tab:radio-ps}
\end{tabular}

$a$: Offset from fit SZ Centroid
\end{table}

\begin{figure}[!h]
\centering
\includegraphics[trim= 0.5in 1.7in 0.5in 2.2in, angle=-90,width=0.28\textwidth]{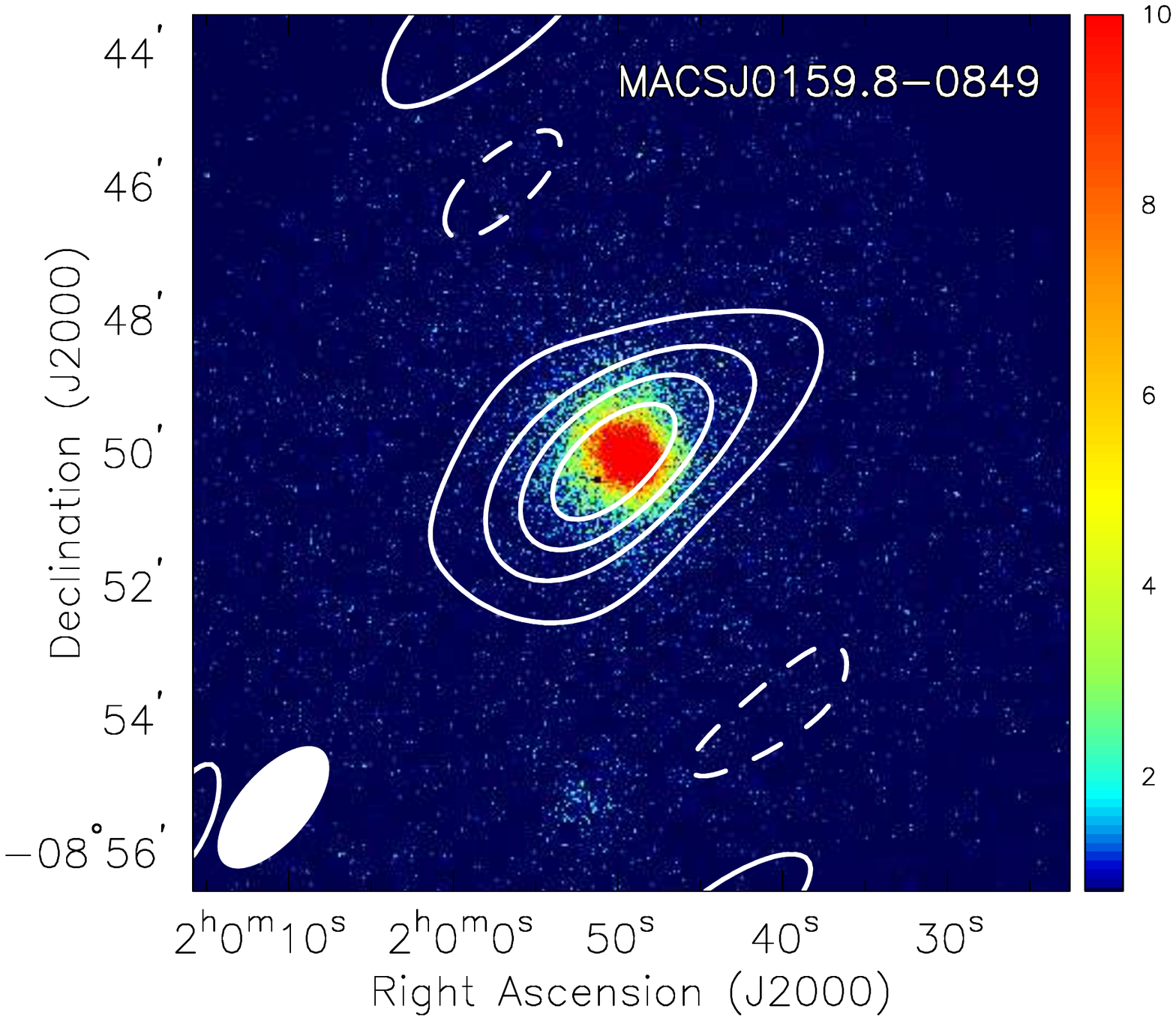}
\includegraphics[trim= 0.5in 1.7in 0.5in 2.2in, angle=-90,width=0.28\textwidth]{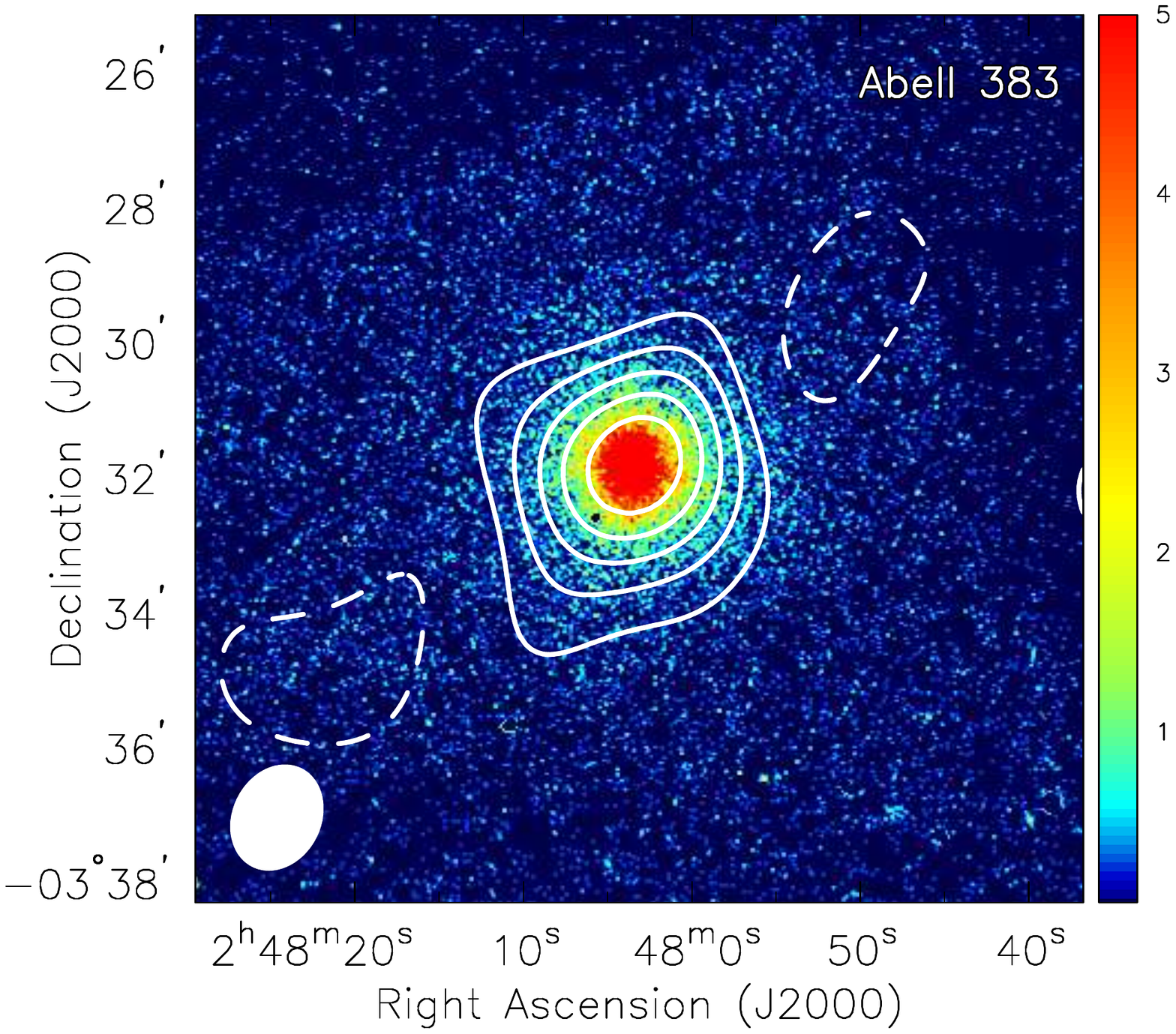}
\includegraphics[trim= 0.5in 1.7in 0.5in 2.2in, angle=-90,width=0.28\textwidth]{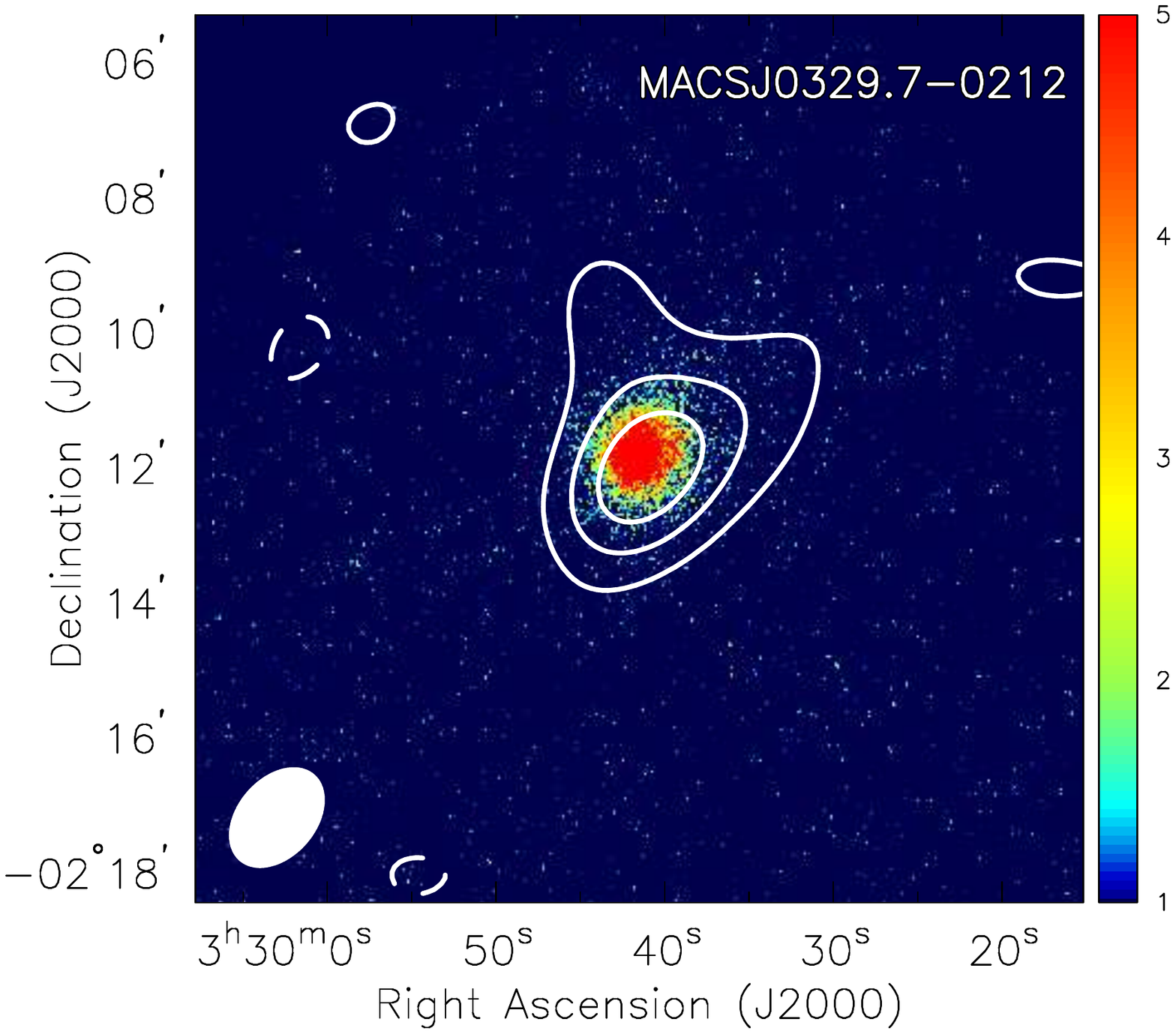}
\includegraphics[trim= 0.5in 1.7in 0.5in 2.2in, angle=-90,width=0.28\textwidth]{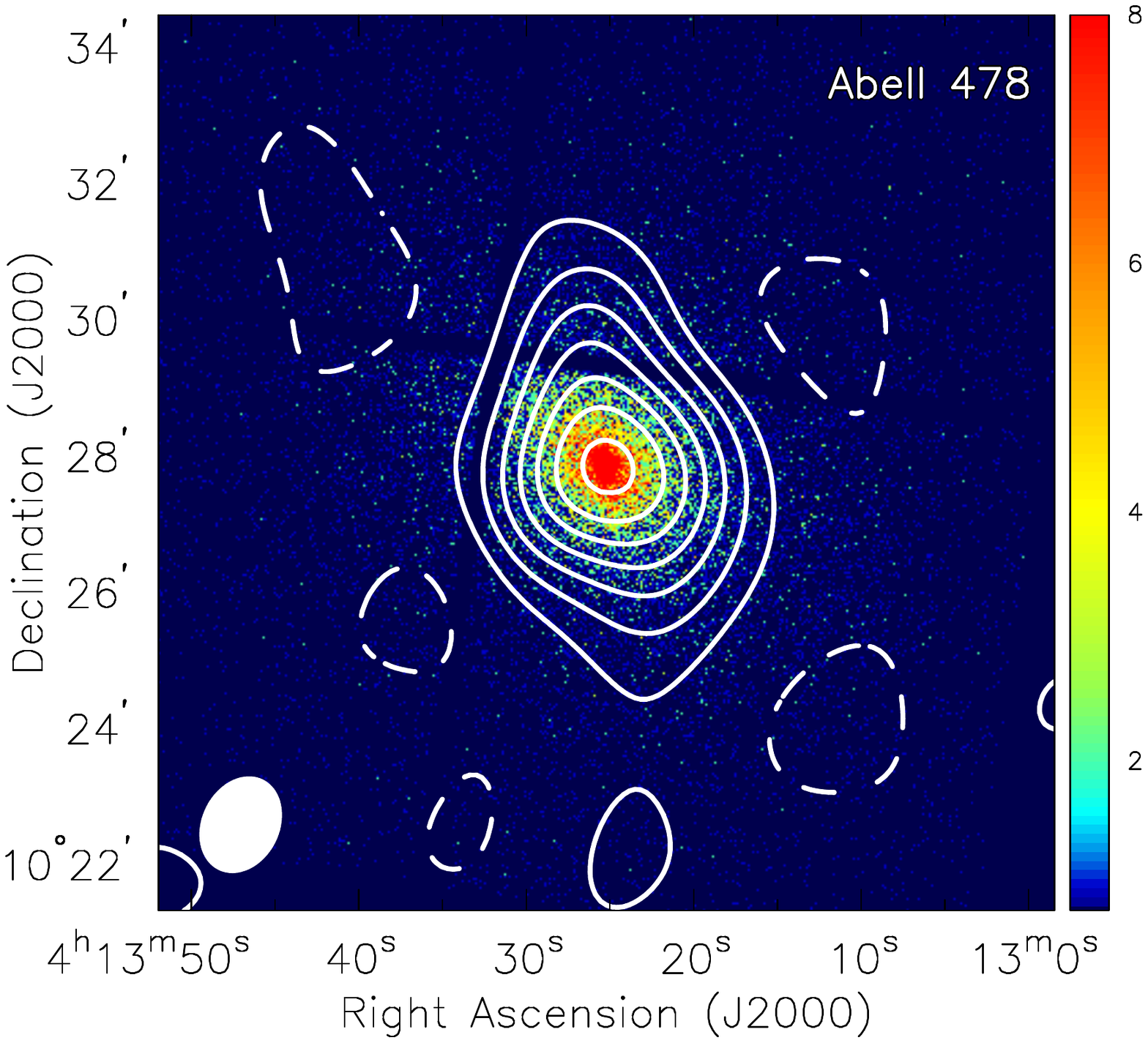}
\includegraphics[trim= 0.5in 1.7in 0.5in 2.2in, angle=-90,width=0.28\textwidth]{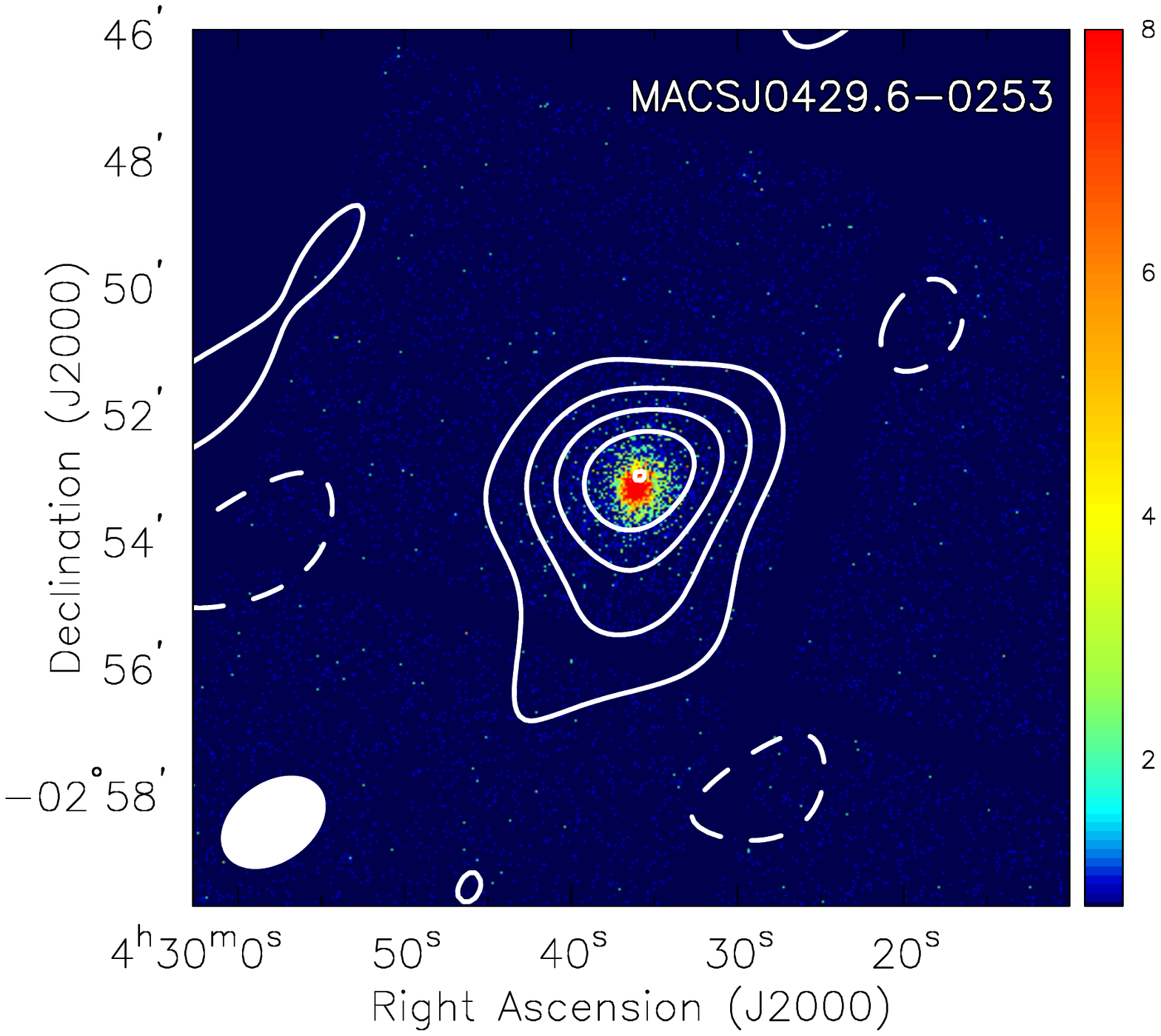}
\includegraphics[trim= 0.5in 1.7in 0.5in 2.2in, angle=-90,width=0.28\textwidth]{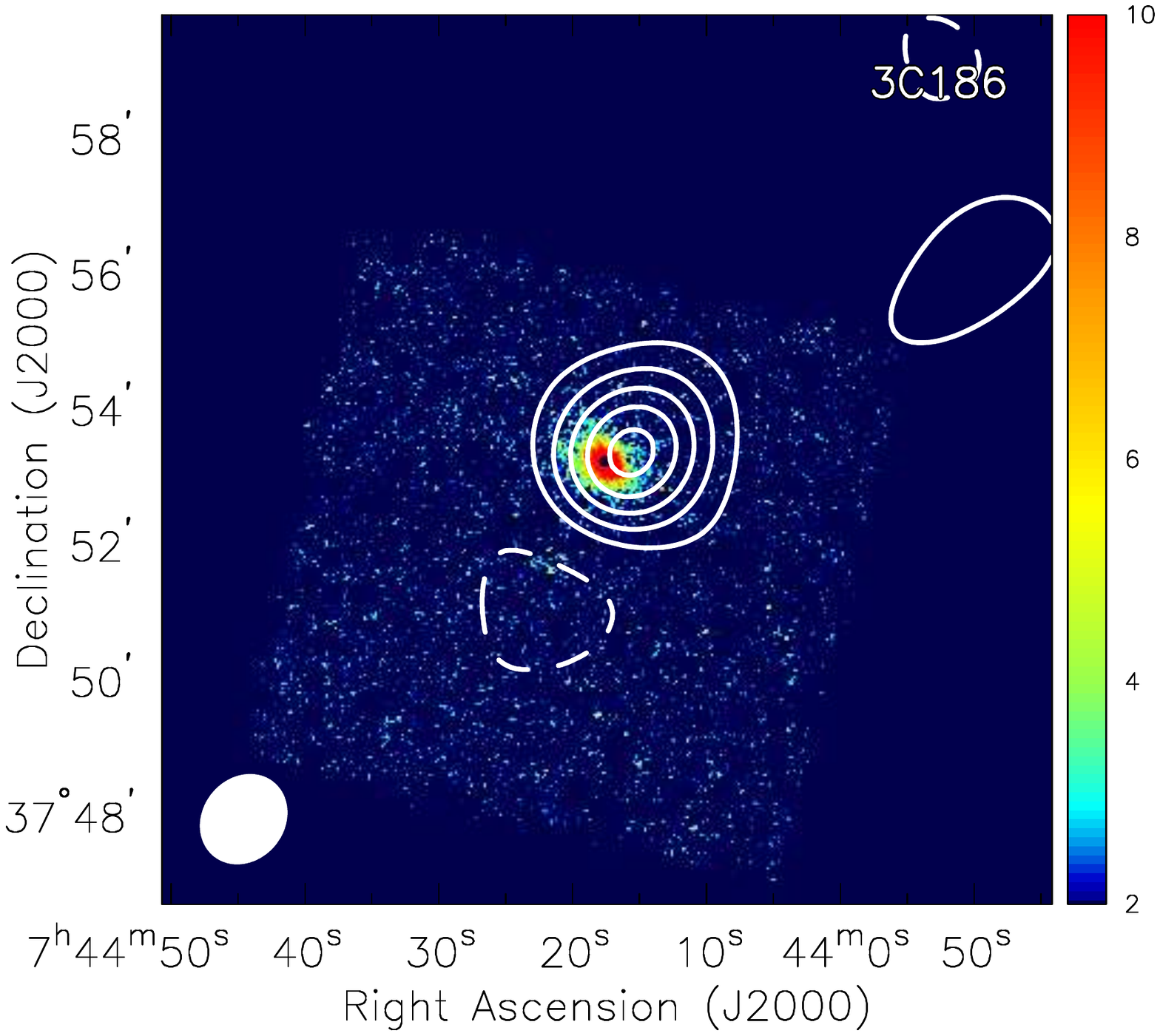}
\includegraphics[trim= 0.5in 1.7in 0.5in 2.2in, angle=-90,width=0.28\textwidth]{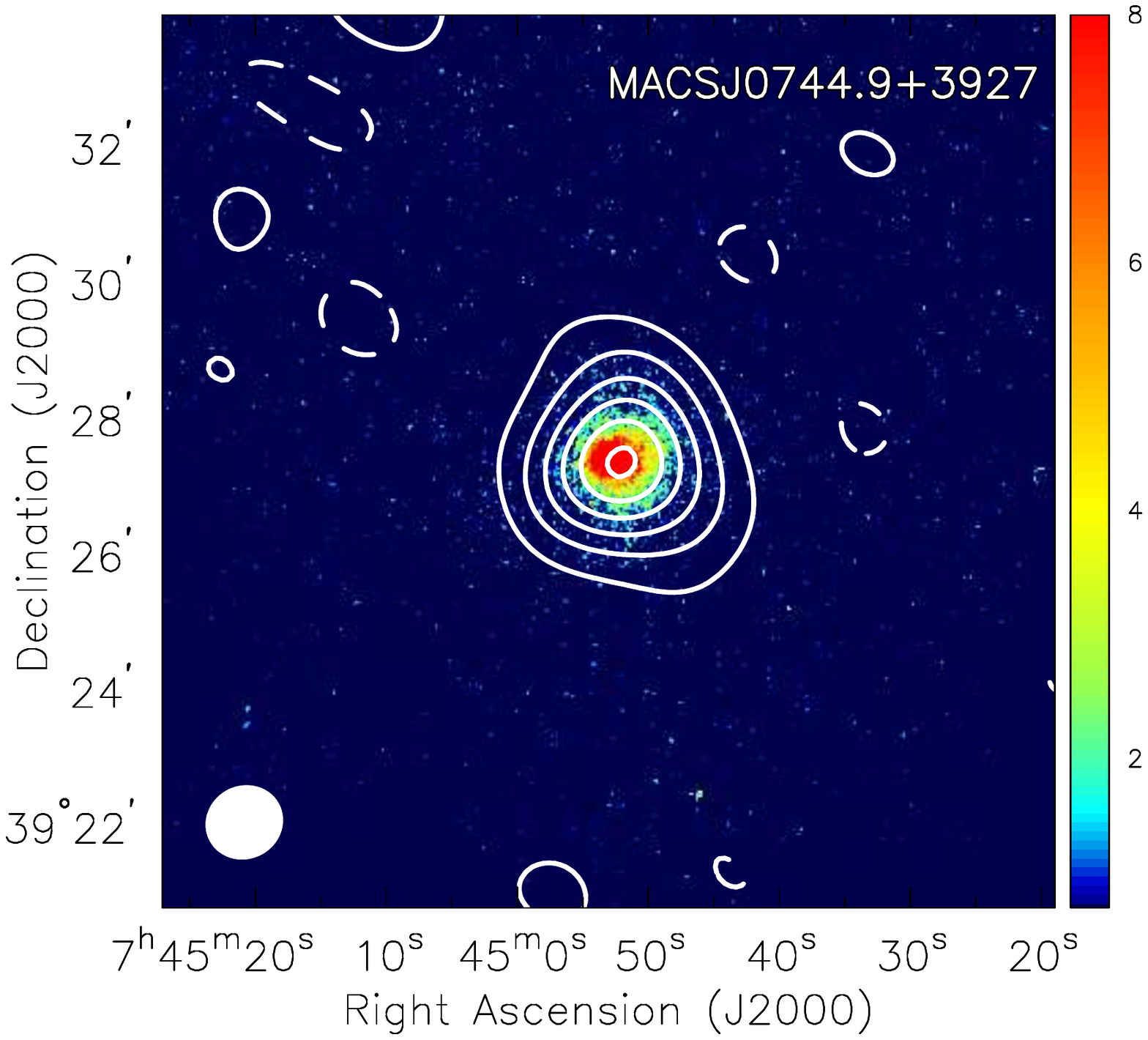}
\includegraphics[trim= 0.5in 1.7in 0.5in 2.2in, angle=-90,width=0.28\textwidth]{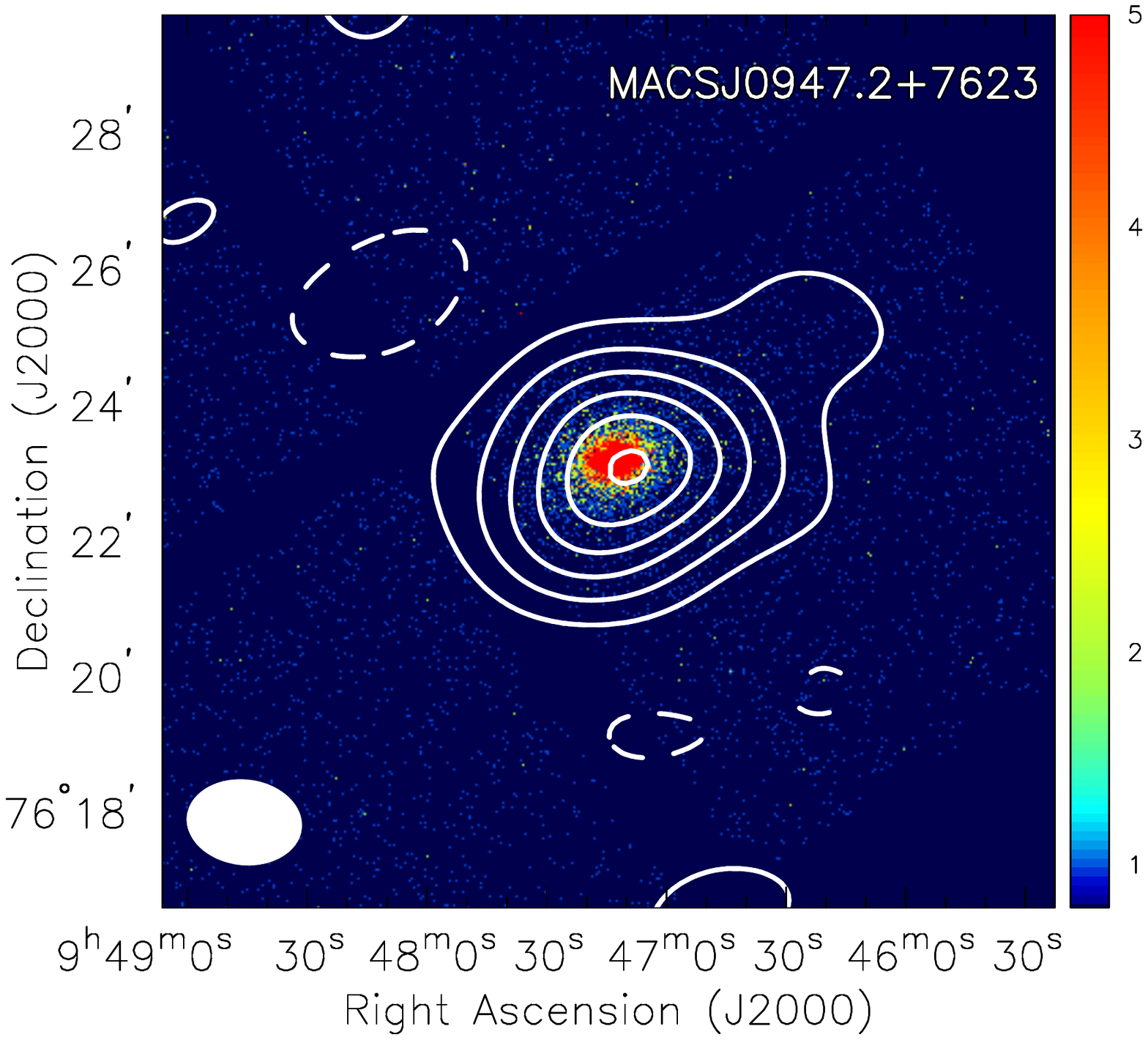}
\includegraphics[trim= 0.5in 1.7in 0.5in 2.2in, angle=-90,width=0.28\textwidth]{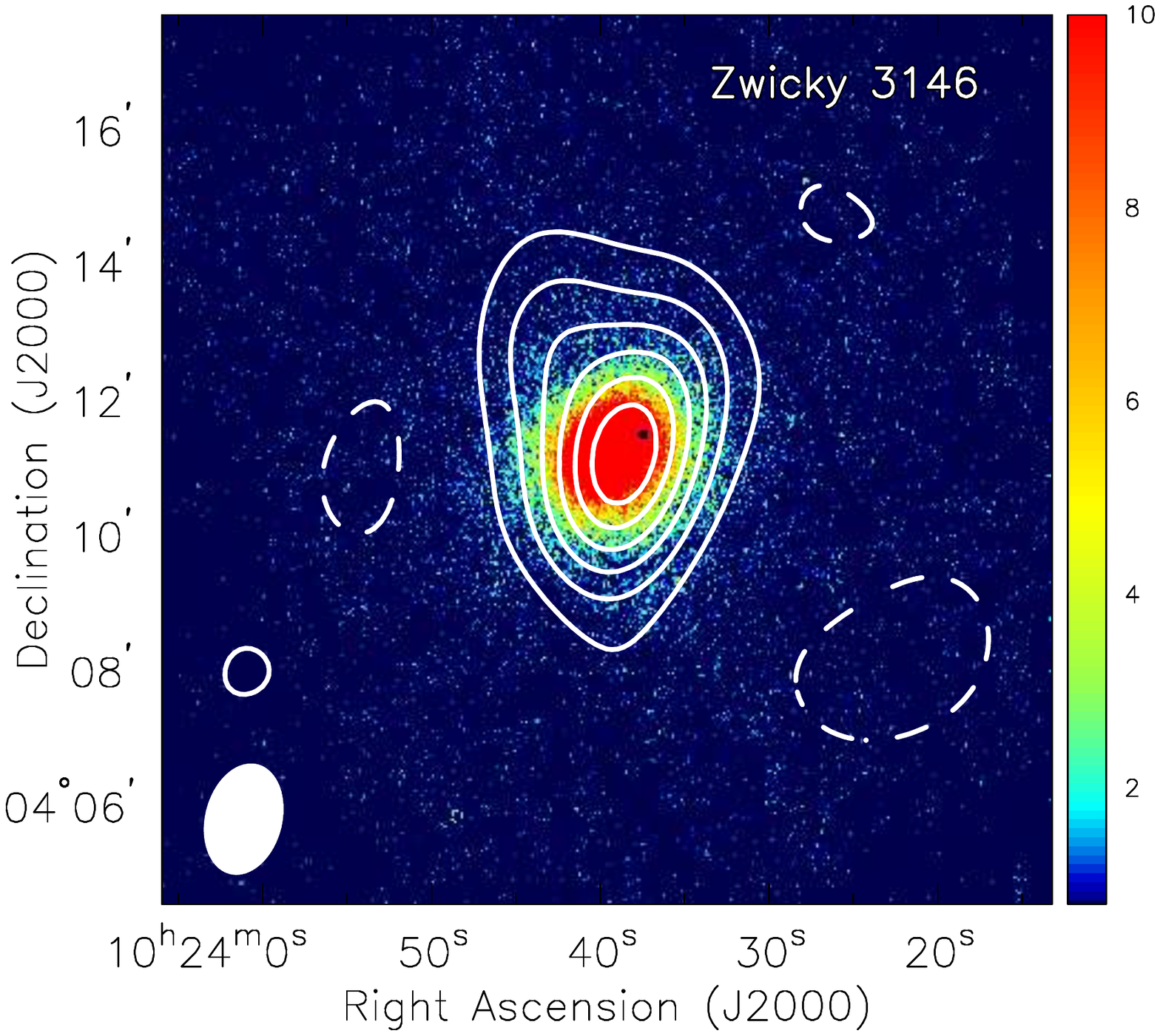}
\includegraphics[trim= 0.5in 1.7in 0.5in 2.2in, angle=-90,width=0.28\textwidth]{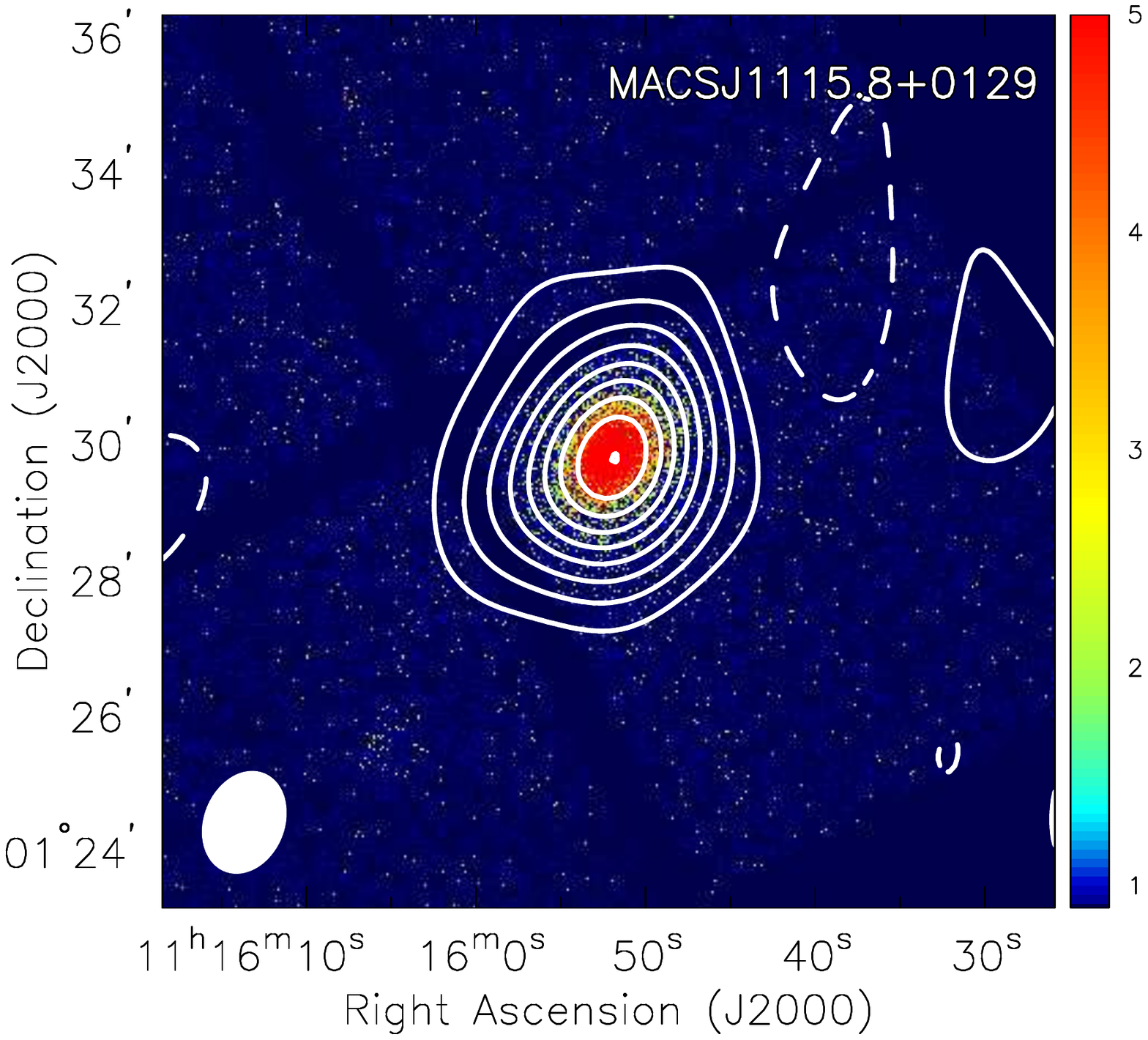}
\includegraphics[trim= 0.5in 1.7in 0.5in 2.2in, angle=-90,width=0.28\textwidth]{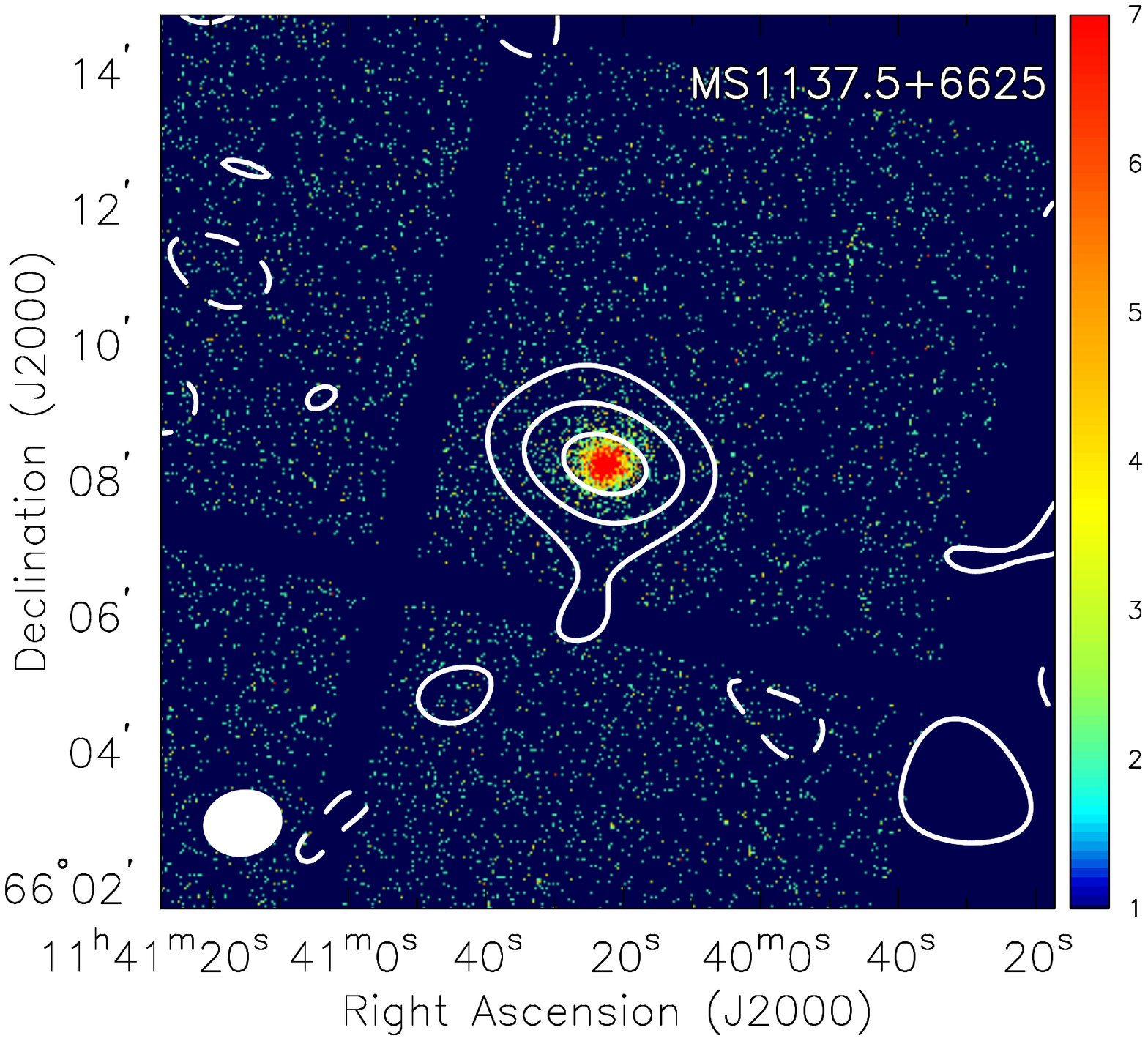}
\includegraphics[trim= 0.5in 1.7in 0.5in 2.2in, angle=-90,width=0.28\textwidth]{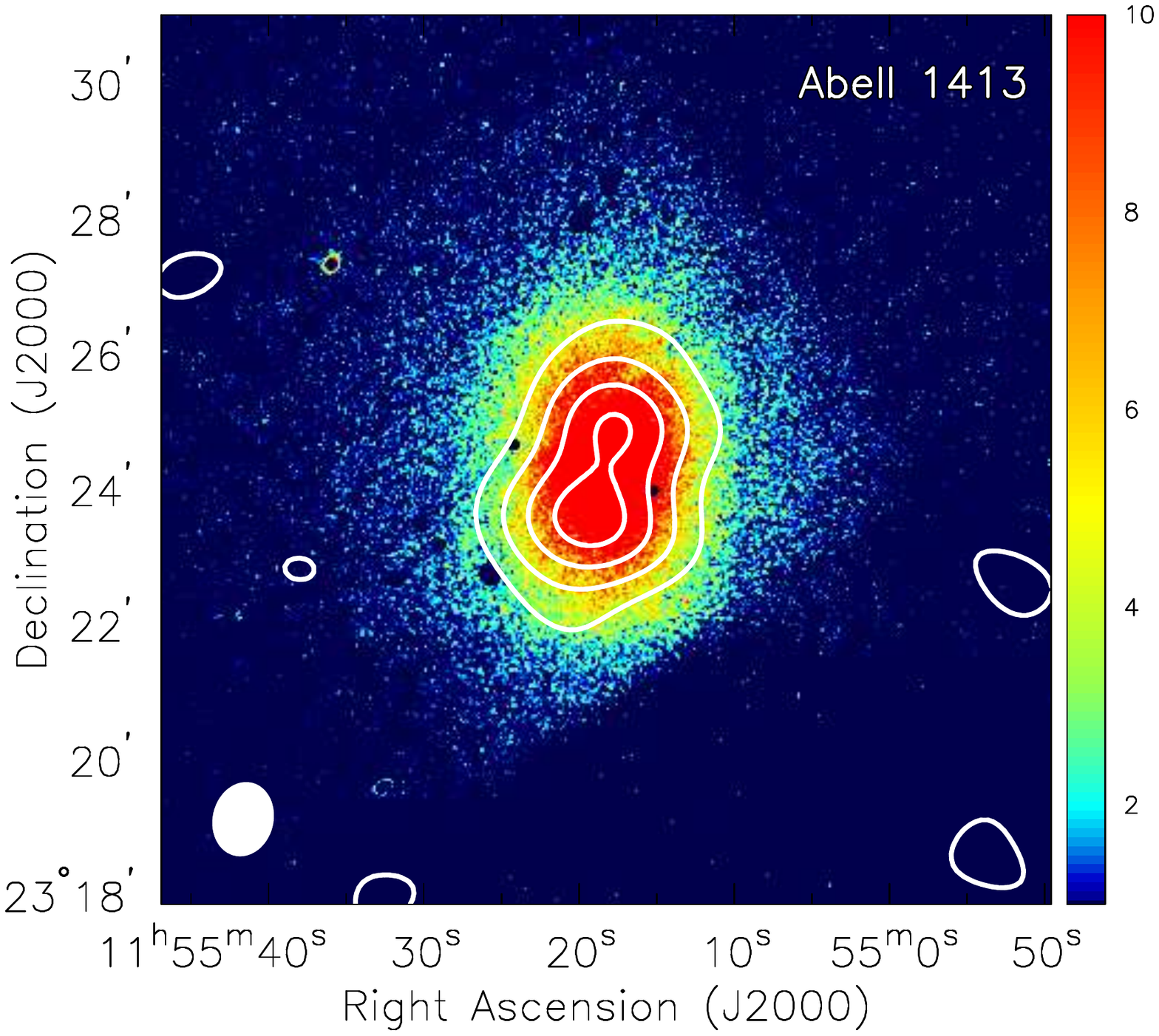}
\includegraphics[trim= 0.5in 1.7in 0.5in 2.2in, angle=-90,width=0.28\textwidth]{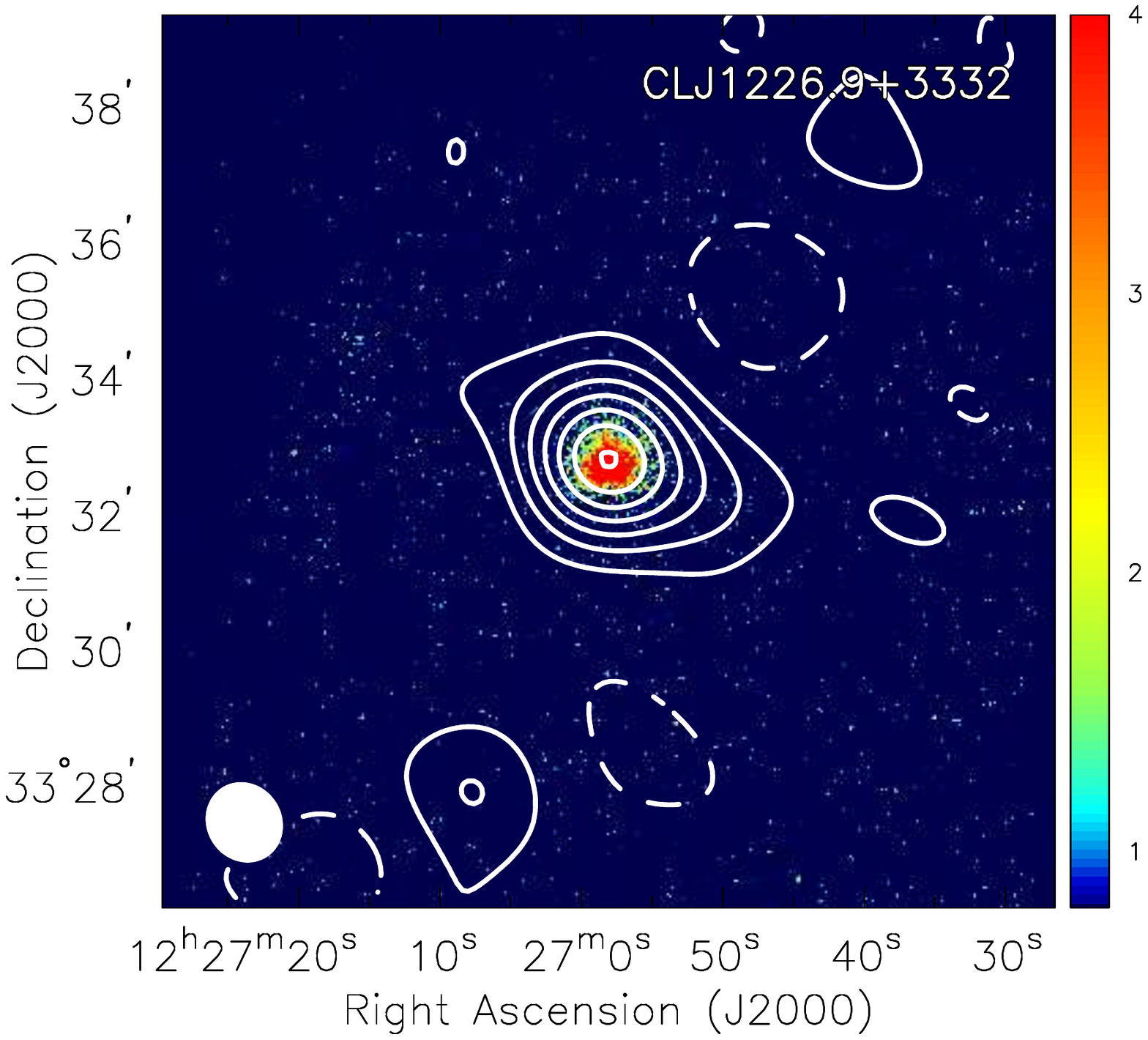}
\includegraphics[trim= 0.5in 1.7in 0.5in 2.2in, angle=-90,width=0.28\textwidth]{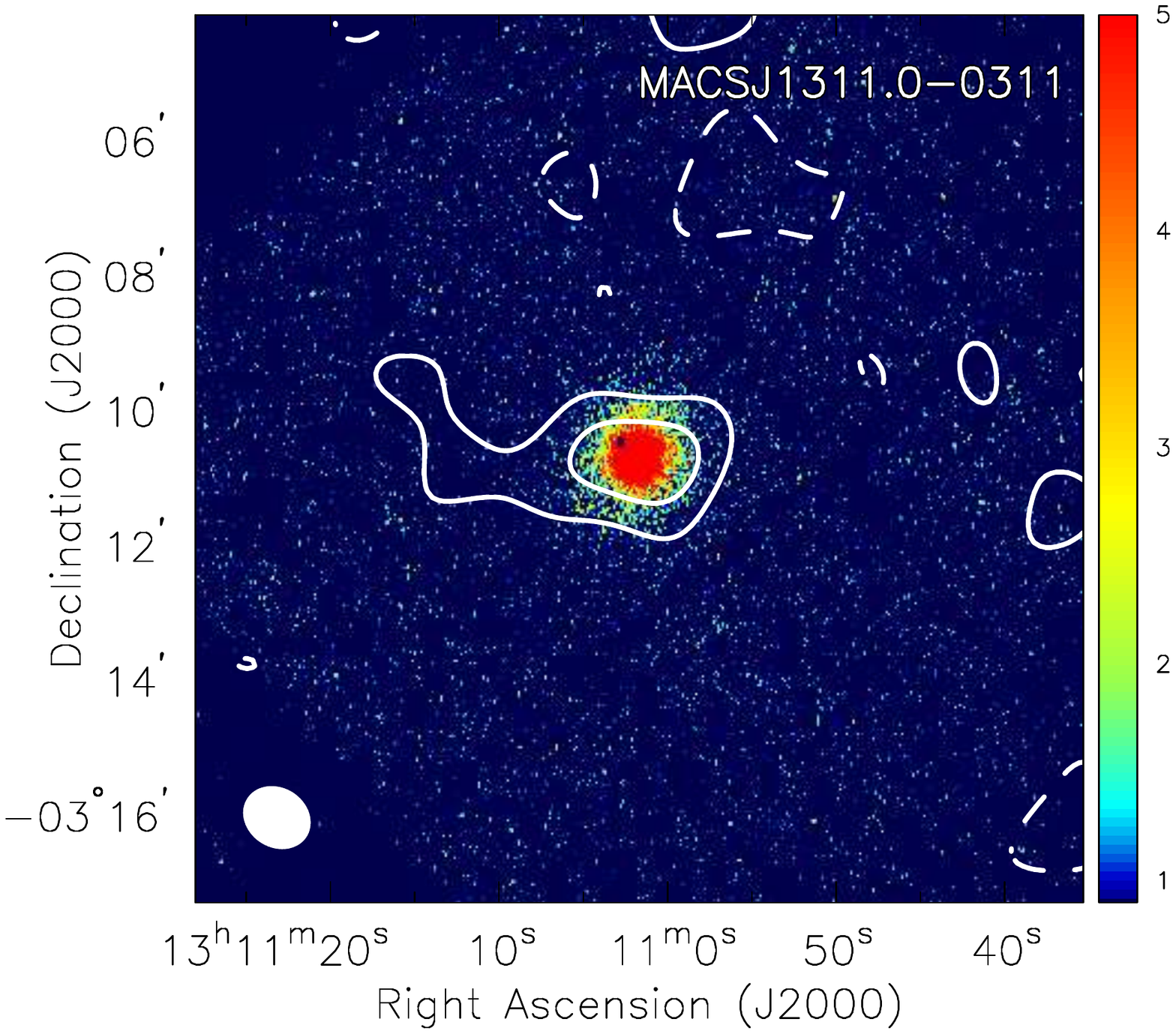}
\includegraphics[trim= 0.5in 1.7in 0.5in 2.2in, angle=-90,width=0.28\textwidth]{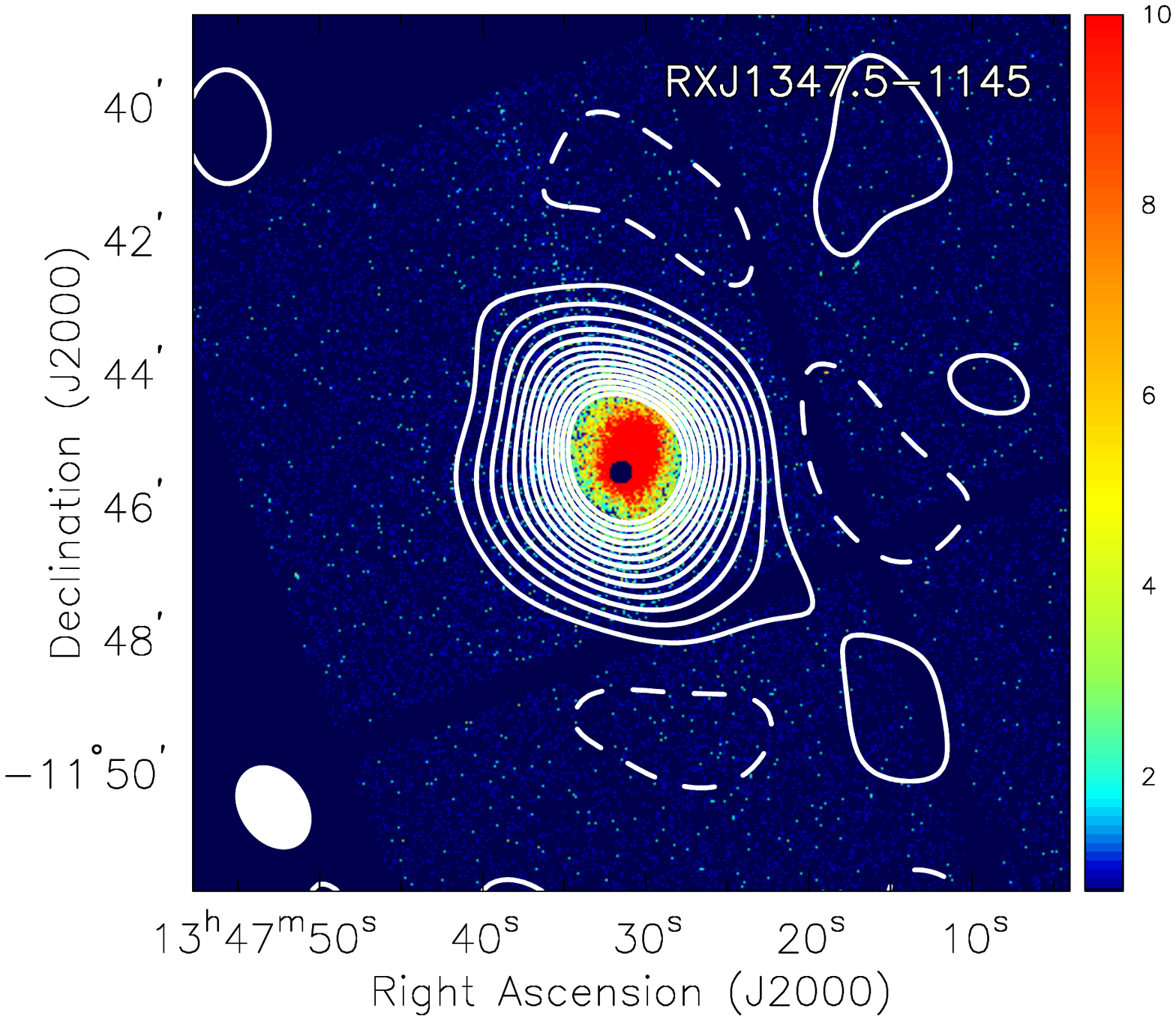}
\end{figure}

\clearpage

\section{Analysis of the \sza\ and \chandra\ data}
\label{sec:analysis}
\subsection{Models for the thermodynamic quantities}
\label{sec:models}
We analyze the SZ and X-ray data using the \cite{bulbul2010} model, which uses
 a consistent parameterization of the electron density, temperature and pressure,
related through the ideal gas law at all radii, i.e., $p_e(r) = n_e(r) kT_e(r)$
for pressure $p_{e}$, electron density $n_{e}$, and temperature $T_e$. 
All thermodynamic quantites depend on the gravitational potential,
\begin{equation}
\phi(r)=\left[ \frac{1}{(\beta-2)} \frac{(1+r/r_s)^{\beta-2} -1}{r/r_s (1+r/r_s)^{\beta-2}} \right],
\end{equation}
in which $\beta$ describes the slope of the matter density at large radii and $r_s$ is a scale radius.
The parameterization of the  \cite{bulbul2010} model does not allow the inner slope
of the matter density to vary,
which is fixed at $r^{-1}$ as in the \cite{navarro1997} model.
The resolution of our SZ data can only effectively constrain the matter
distribution on scales larger than the synthesized beam, which is of order 
1 arcmin for these observations, and therefore we would not be able to place
significant constraints on the inner slope.
As explained in \cite{bulbul2010}, the potential is continuous at $\beta=2$, the value
of the \cite{navarro1997} mass density model.
The radial electron temperature profile is given by
\begin{equation}
T_e(r) = T_0 \phi(r)   \tau_{cool}(r),
\end{equation}
where $\tau_{cool}(r)$ is the \cite{vikhlinin2006} phenomenological
core taper function, required to fit cool-core clusters, which is equal to one at large radii.
The density is parameterized as 
\begin{equation}
n_e(r) = n_{e0} \phi(r)^n \tau_{cool}^{-1}(r).
\end{equation}
in such a way that the pressure distribution is not altered by the presence of the cool core.
At large radii, where the effect of the cool core vanishes, the thermodynamic quantities
are related by a simple polytropic equation of state.
The electron pressure profile is therefore parameterized as
\begin{equation}
p_e(r) = P_{e0} \phi(r)^{n+1}
\label{eq:pressure}
\end{equation}
and is independent of the presence of a cool core. 
The model therefore has five independent parameters for non-cool-core clusters: the scale radius $r_s$;
the index $\beta$, the polytropic index $n$, and the normalization constants
for the three thermodynamic quantities  which satisfy $n_{e0} k T_0=P_{e0}$.
For cool-core clusters, the $\tau_{cool}$ function 
\begin{equation}
\tau_{cool}(r)=\frac{\alpha+(r/r_{cool})^{\gamma}}{1+(r/r_{cool})^{\gamma}}
\end{equation}
adds three additional adjustable parameters.

To test for model-dependent biases, we also use the \cite{arnaud2010} model
to fit the SZ data.  This model describes the cluster pressure profile using an
analytic function motivated by numerical simulations \citep{nagai2007b}
and X-ray observations of the REXCESS sample,
\begin{equation}
p_e(r) = \frac{p_{e,i}}{(r/r_p)^c \left[1+(r/r_p)^a\right]^{(b-c)/a}}.
\label{eq:a10press}
\end{equation}
The parameters $p_{e,i}$ and $r_p$ are left free in our fits 
to the SZ effect observations.
The values $(a,b,c)$ are the power law indices that describe the (intermediate, outer, 
inner) slopes of $p_e(r)$.  We use the ``universal'' values 
$(a,b,c) = (1.05, 5.49, 0.31)$ obtained by \cite{arnaud2010} 
from a fit to X-ray observations of the REXCESS sample.  Note that 
\citet{arnaud2010} find different best-fit values for cool-core clusters.
We choose to use the parameters fit to the entire sample because 
our sample was not selected based on the presence of a cool core, 
and in fact contains a few non cool-core clusters, namely 
3C186, MS1137.5+6625 and CLJ1226.9+3332.

\subsection{Method of analysis}
\label{sec:method}

As in previous work with the SZA \citep[e.g.,][]{mroczkowski2009,hasler2011}, 
we relate the point-source subtracted interferometric 
SZ visibilities to the unitless integrated Compton $y$ by introducing $Y(u,v)$, defined as
\begin{equation}
\label{eq:Yuv}
Y(u,v) \equiv \frac{V_\nu(u,v)}{g(x) \, I_0}.
\end{equation}
Here $g(x)$ corrects for the frequency dependence of the SZ flux, and
$I_0 = 2 (k_B T_{CMB})^3/(h c)^2$ is the primary CMB intensity.
The SZ models and compact radio sources are fit directly and simultaneously in Fourier space, 
where the statistical properties of the model fits are better understood and the noise is Gaussian. 
This is done simply by building up the sky brightness image, Fourier transforming it, and computing 
the likelihood of the model.

The X-ray data consist of spectroscopic temperature measurements taken in cluster-centric annuli, 
and an X-ray image in units of surface brightness (counts s$^{-1}$ cm$^{-2}$ sr$^{-1}$).
The X-ray surface brightness $S_x$ varies with the line of sight integral of the electron 
density and temperature distributions as
\begin{equation}
S_x = \frac{1}{4 \pi (1+z)^3} \int n_{e}^{2} \Lambda_{ee}(T_{e}, A) d\ell,
\label{eqn:sx}
\end{equation}
where $\ell$ is the line of sight through the cluster,
$n_{e}$ is the electron density, $T_e$ is the electron temperature,
$A$ is the metallicity, and $\Lambda_{ee}(T_e,A)$ is the X-ray cooling function
(in units of counts cm$^3$ s$^{-1}$) as a function of
electron temperature and metallicity. Each cluster was divided in
a number of annuli according to the total number of photons detected, and 
for each annular region the temperature and
abundance were free parameters.
The surface brightness is only marginally sensitive to the choice of outer 
limit of integration in Equation~\ref{eqn:sx}: we find that the masses vary by less than 1\% 
when the outer limit ranges between 2 and 5 Mpc.  
We therefore choose a limit of 2 Mpc, which corresponds to approximately 
the virial radius for clusters in this mass range.
We use the \cite{mazzotta2004} definition of spectroscopic temperature in the
comparison of model and observed temperatures in each annulus.

We first estimate the pressure profile of the ICM by jointly fitting the
SZ and X-ray data with the \cite{bulbul2010} model.
Both datasets are used simultaneously to constrain all three thermodynamic
quantities, with the global shape parameters $\beta$, $n$ and $r_s$ (and the
cool-core parameters when applicable) linked among the profiles.
Both datasets contribute to the determination of the shape of the pressure profile, with SZ observations
contributing primarily at the largest radii where the sensitivity of \chandra\ to the diffuse cluster emission
is limited. 
Instead of linking the normalization of the pressure profile ($P_{e0}$)
to the product of the normalizations of the density and temperature ($n_{e0}$ and $T_{0}$),  
we let the normalizations be free, and check \emph{a posteriori} that $P_{e0}=n_{e0} \times k T_{0}$
in accordance with the ideal gas law.  The normalization
of the pressure is determined by the  SZ data, and the normalizations of temperature and density by the X-ray data.

This method results in the measurement of the shape of the pressure profile, $p_e(r)/P_{e0}$,
and two normalizations determined independently by each of the two datasets.
The two normalizations are left free to vary because in principle systematic uncertainties
in the two datasets could lead to different values, and we do not want to assume an \emph{a priori}
agreement between them.
The fit uses a Markov chain Monte Carlo method \citep{bonamente2004}, and computes
the angular diameter distance assuming a 
$\Omega_{\Lambda}=0.73$, $\Omega_M=0.27$ and $h=0.73$ cosmology.

To obtain a measurement of the integrated pressure that depends only on the SZ data, we also
perform another fit in which we fix the shape parameters of the \cite{bulbul2010} pressure profile
to $n=3.5$ and $\beta=2.0$. These values correspond to the median of the values
obtained from the joint fit. This pressure profile with fixed slope parameters is
directly comparable to the universal pressure profile of \cite{arnaud2010}, since both are determined
by modelling of high-resolution X-ray data (from fits to the REFLEX sample for the \citealt{arnaud2010} model,
and from fits to the \citealt{allen2008} observations for our model), and have just two free parameters
(scale radius and normalization constant).  In the following, we refer to this 2-parameter model
as the \cite{bulbul2010} average pressure profile.

Measurements of the ICM pressure using SZ and X-ray data
are subject to different sources of systematic uncertainty that could
affect the calculation of the $Y_{\rm sph}$ parameter
\citep{hasler2011}.
Systematic errors that integrate down with sample size include
cluster asphericity, the effect of X-ray background, and the presence of kinetic SZ effect;
these errors are included in the calculation
of the ratio between the various measurements of $Y_{\rm sph}(r_{500})$,
and of the weighted averages and $\chi^2_{min}$ values in Sections~\ref{sec:joint}
and \ref{sec:joint-vs-universal}, following the prescriptions of \cite{hasler2011}.

\section{Integrated pressure measurements}
\label{sec:Y}
\subsection{Joint SZ and X-ray fit using the \cite{bulbul2010} model}
\label{sec:joint}

The integrated pressure, which we quantify in terms of the Compton $y$ parameter,
is expected to be a good proxy for total cluster mass.
Since the \sza\ measures the integrated flux within Fourier modes on the sky, 
our SZ data relate most directly to the integrated Compton-$y$ parameter $Y_{\rm cyl}$. 
However, it is conventional in X-ray analyses to report spherically integrated
quantities.  We therefore quantify the integrated pressure using the spherically-integrated 
Compton $y$ parameter $Y_{\rm sph}$ out to $r_{500}$.  The overdensity radius $r_{500}$ is given by
\begin{equation}
r_{\Delta} = \left(\frac{M_{tot}(r_{\Delta})}{\frac{4}{3}\pi \cdot \Delta \rho_c(z)} \right)^{1/3}
\end{equation}
with $\Delta=500$, where $\rho_c(z)$ is the critical density of the universe at the
cluster redshift.  The total cluster mass is calculated under the assumption of
hydrostatic equilibrium; for the \cite{bulbul2010} model,
the total mass is given by
\begin{equation}
M_{tot}(r)= \frac{4\pi\rho_{i}r_{s}^{3}}{(\beta-2)}\left( \frac{1}{\beta-1} +\frac{1/(1-\beta) - r/r_s}{(1+r/r_{s})^{\beta-1}}\right)\tau_{cool}(r),
\label{eqn:mtot}
\end{equation}
where the matter density normalization is given by 
$\rho_{i}=(k T_{0} (n+1) (\beta-1))/(4 \pi G \mu m_{p} r_{s}^{2})$;
$\mu$ is the mean molecular weight, and $m_p$ is the proton mass.

Using the method of analysis discussed in Section~\ref{sec:method},
we first compare $Y_{\rm sph}$ normalized using $n_{e0}$ and $T_0$ 
constrained by the X-ray data with $Y_{\rm sph}$ normalized
using $P_{e0}$ constrained by the SZ data.
This comparison is summarized in Table~\ref{tab:pres-y}.
The normalizations are in good agreement: the weighted
average of the ratio between the measurements using the SZ and
X-ray normalization is 
$1.06\pm0.04$.  
This indicates that systematic uncertainties do not produce a large
overall offset between the two observables.

Below, we refer to $Y_{\rm sph}$ as the measurement obtained from the joint fit using the
X-ray normalization. We adopt this value since the joint profile makes use
of all information available from both the X-ray and SZ observations including the 
effect of the cool core, and since both normalizations are in agreement.

\begin{table}[h]
\centering
\caption{Measurement of Integrated $Y_{\rm sph}$ at $r_{500}$ from Joint X-ray and SZ Data Using the Polytropic Model}
\scriptsize
\begin{tabular}{l|ccc|c}
\hline \hline
Cluster & $r_{500}$ 		& \multicolumn{2}{c}{$Y_{\rm sph} (r_{500})$} & 
                  \\
        &                       & SZ normalization  & X-ray normalization & SZ-toX-ray ratio\\
	& ($^{\prime\prime}$) 	& $(10^{-11})$ 	   & $(10^{-11})$    & \\
\hline
\\[-0.6pc]
 MACSJ0159.8-0849 & $221.1\pm^{11.0}_{12.3}$ & $8.30\pm^{0.76}_{0.88}$ & $9.67\pm^{1.14}_{1.12}$ & $0.86\pm^{0.08}_{0.07}$\\[0.2pc]
 Abell~383        & $268.5\pm^{22.1}_{20.7}$ & $4.92\pm^{0.77}_{0.70}$ & $4.19\pm^{0.82}_{0.70}$ & $1.17\pm^{0.16}_{0.15}$ \\[0.2pc] 
 MACS0329.7-0212  & $138.4\pm^{12.7}_{11.9}$ & $3.03\pm^{0.53}_{0.49}$ & $2.61\pm^{0.59}_{0.49}$ & $1.16\pm^{0.20}_{0.18}$ \\[0.2pc]
 Abell~478        & $714.3\pm^{23.5}_{34.5}$ & $49.61\pm^{3.15}_{3.19}$ & $60.62\pm^{4.84}_{6.13}$ & $0.82\pm^{0.06}_{0.05}$\\[0.2pc]
 MACSJ0429.6-0253 & $182.3\pm^{18.5}_{15.1}$ & $2.75\pm^{0.49}_{0.43}$ & $3.30\pm^{0.81}_{0.61}$ & $0.83\pm^{0.14}_{0.14}$ \\[0.2pc]
 3C186            & $72.1\pm^{5.5}_{5.7}$    & $1.01\pm^{0.23}_{0.20}$ & $0.86\pm^{0.14}_{0.13}$ & $1.17\pm^{0.24}_{0.21}$\\[0.2pc]
 MACSJ0744.9+3927 & $120.3\pm^{8.6}_{7.5}$   & $5.04\pm^{0.66}_{0.57}$ & $3.80\pm^{0.72}_{0.57}$ & $1.33\pm^{0.16}_{0.15}$ \\[0.2pc]
 MACSJ0947.2+7623 & $196.2\pm^{15.1}_{15.5}$ & $5.18\pm^{0.73}_{0.70}$ & $6.00\pm^{1.17}_{1.11}$ & $0.86\pm^{0.14}_{0.11}$ \\[0.2pc]
 Zwicky~3146      & $265.7\pm^{8.7}_{8.6}$   & $12.17\pm^{1.22}_{1.17}$ & $10.56\pm^{0.93}_{0.95}$ & $1.14\pm^{0.13}_{0.10}$\\[0.2pc]
 MACSJ1115.8+0129 & $200.0\pm^{9.6}_{10.7}$  & $7.74\pm^{0.57}_{0.58}$ & $6.26\pm^{0.78}_{0.83}$ & $1.24\pm^{0.14}_{0.11}$\\[0.2pc]
 MS1137.5+6625    & $78.8\pm^{5.6}_{5.1}$    & $1.08\pm^{0.16}_{0.15}$ & $0.73\pm^{0.11}_{0.10}$ & $1.49\pm^{0.26}_{0.24}$\\[0.2pc]
 Abell~1413       & $454.4\pm^{20.3}_{20.3}$ & $17.10\pm^{2.64}_{2.03}$ & $23.12\pm^{2.36}_{2.32}$ & $0.75\pm^{0.07}_{0.06}$ \\[0.2pc]
 CLJ1226.9+3332   & $109.4\pm^{8.3}_{8.3}$   & $3.31\pm^{0.34}_{0.34}$ & $3.06\pm^{0.54}_{0.52}$ & $1.09\pm^{0.18}_{0.15}$\\[0.2pc]
 MACSJ1311.0-0311 & $156.5\pm^{11.5}_{10.2}$ & $2.36\pm^{0.61}_{0.58}$ & $2.31\pm^{0.36}_{0.31}$ & $1.02\pm^{0.26}_{0.23}$\\[0.2pc]
 RXJ1347.5-1145   & $218.0\pm^{6.6}_{5.9}$   & $14.02\pm^{0.75}_{0.75}$ & $21.59\pm^{1.82}_{1.82}$ & $0.65\pm^{0.04}_{0.04}$\\[0.2pc]
 Abell~1835       & $370.7\pm^{7.6}_{8.0}$   & $31.41\pm^{1.56}_{1.56}$ & $29.67\pm^{1.56}_{1.57}$ & $1.06\pm^{0.06}_{0.06}$\\[0.2pc]
 MACSJ1423.8+2404 & $189.1\pm^{16.4}_{15.4}$ & $2.15\pm^{0.45}_{0.39}$ & $2.52\pm^{0.57}_{0.51}$ & $0.86\pm^{0.22}_{0.18}$\\[0.2pc]
 MACSJ1427.3+4408 & $150.5\pm^{4.2}_{4.6}$   & $3.39\pm^{0.57}_{0.50}$ & $4.75\pm^{0.44}_{0.46}$ & $0.72\pm^{0.11}_{0.11}$\\[0.2pc]
 RXJ1504.1-0248   & $326.7\pm^{12.1}_{9.9}$  & $15.73\pm^{1.43}_{1.30}$ & $18.03\pm^{1.54}_{1.26}$ & $0.87\pm^{0.06}_{0.06}$\\[0.2pc]
 MACSJ1532.9+302  & $189.1\pm^{9.9}_{9.1}$   & $5.04\pm^{0.65}_{0.55}$ & $4.61\pm^{0.61}_{0.50}$ & $1.09\pm^{0.12}_{0.12}$\\[0.2pc]
 MACSJ1621.6+3810 & $147.7\pm^{8.0}_{11.1}$  & $2.53\pm^{0.29}_{0.30}$ & $2.76\pm^{0.37}_{0.48}$ & $0.93\pm^{0.12}_{0.10}$\\[0.2pc]
 Abell~2204       & $504.6\pm^{12.5}_{11.2}$ & $44.97\pm^{2.99}_{2.74}$ & $43.93\pm^{3.08}_{2.59}$ & $1.02\pm^{0.05}_{0.05}$ \\[0.2pc]
 MACSJ1720.3+3536 & $170.5\pm^{8.6}_{8.1}$   & $3.89\pm^{0.30}_{0.29}$ & $3.93\pm^{0.49}_{0.42}$ & $0.98\pm^{0.10}_{0.09}$\\[0.2pc]
 RXJ2129.6+0005   & $297.6\pm^{13.1}_{13.6}$ & $10.78\pm^{1.04}_{1.02}$ & $10.48\pm^{1.34}_{1.22}$ & $1.04\pm^{0.10}_{0.11}$\\[0.2pc]
 Abell~2537       & $256.2\pm^{13.4}_{14.4}$ & $7.27\pm^{0.81}_{0.77}$ & $7.37\pm^{0.97}_{0.95}$ & $0.99\pm^{0.09}_{0.09}$\\[0.2pc]
\hline
\label{tab:pres-y}
\end{tabular}
\end{table}

\subsection{SZ-only fit using the \cite{bulbul2010} average pressure profile}
\label{sec:joint-vs-universal}

\begin{figure}[!t]
\centering
\includegraphics[width=3.5in,angle=-90]{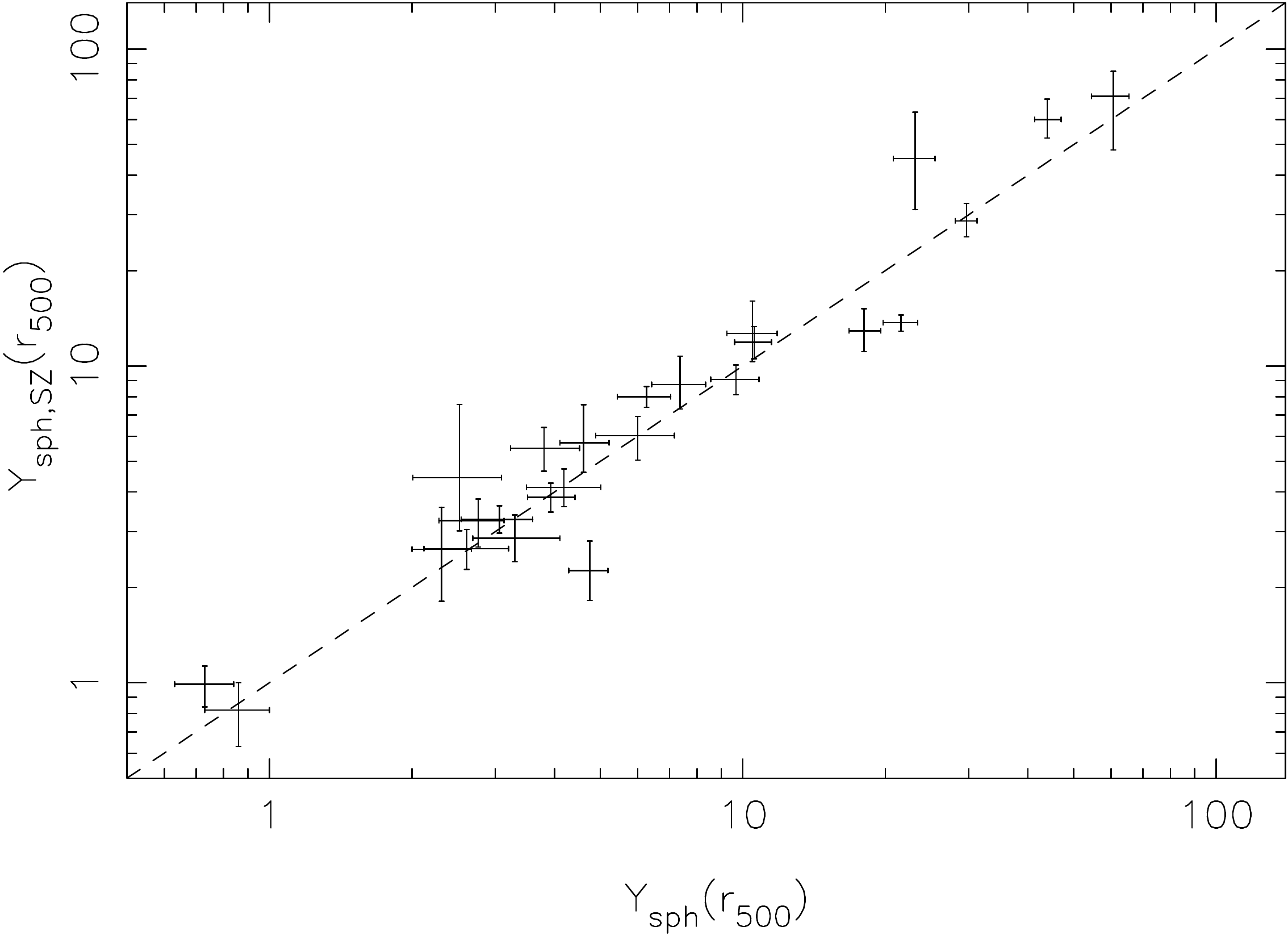}
\caption{Integrated pressure ($Y_{\rm sph, SZ, B10}$) from SZ data plotted against 
integrated  pressure ($Y_{\rm sph}$) from the joint analysis, both measured out to
the same value of $r_{500}$.
 The dashed line is the curve $y=x$.}
\label{fig:y}
\end{figure}

We also fit only the \sza\ data to the \cite{bulbul2010} average pressure profile,
which consists of the pressure profile of Equation~\ref{eq:pressure} 
with $P_{e0}$ and $r_s$ as free parameters and the two shape parameters
fixed at $n=3.5$ and $\beta=2.0$. 
We use this model to compute $Y$ as described above, which we refer to as $Y_{\rm sph, SZ, B10}$.
The value of $r_{500}$ used in computing $Y_{\rm sph, SZ, B10}$ is determined 
from the joint fit.  These results are shown in Table~\ref{tab:pres-params-b10},
and are plotted against the joint fit $Y_{\rm sph}$ in Figure~\ref{fig:y}.
We find that the weighted mean of the ratio between the measurements
is given by $Y_{\rm sph, SZ, B10}/Y_{\rm sph}=0.90\pm0.05$, where the uncertainty is
the standard deviation of the weighted mean.
A linear fit of the two measurements to a $y=x$ model results in a $\chi^2_{min}=35.3$ 
for 25 degrees of freedom, and we measure a scatter of 16\%.

\subsection{Comparison between the  \cite{bulbul2010} and  \cite{arnaud2010} pressure profiles
applied to the SZ data}
\label{sec:b10-vs-a10}

\begin{figure}[!t]
\centering
\includegraphics[width=3.5in,angle=-90]{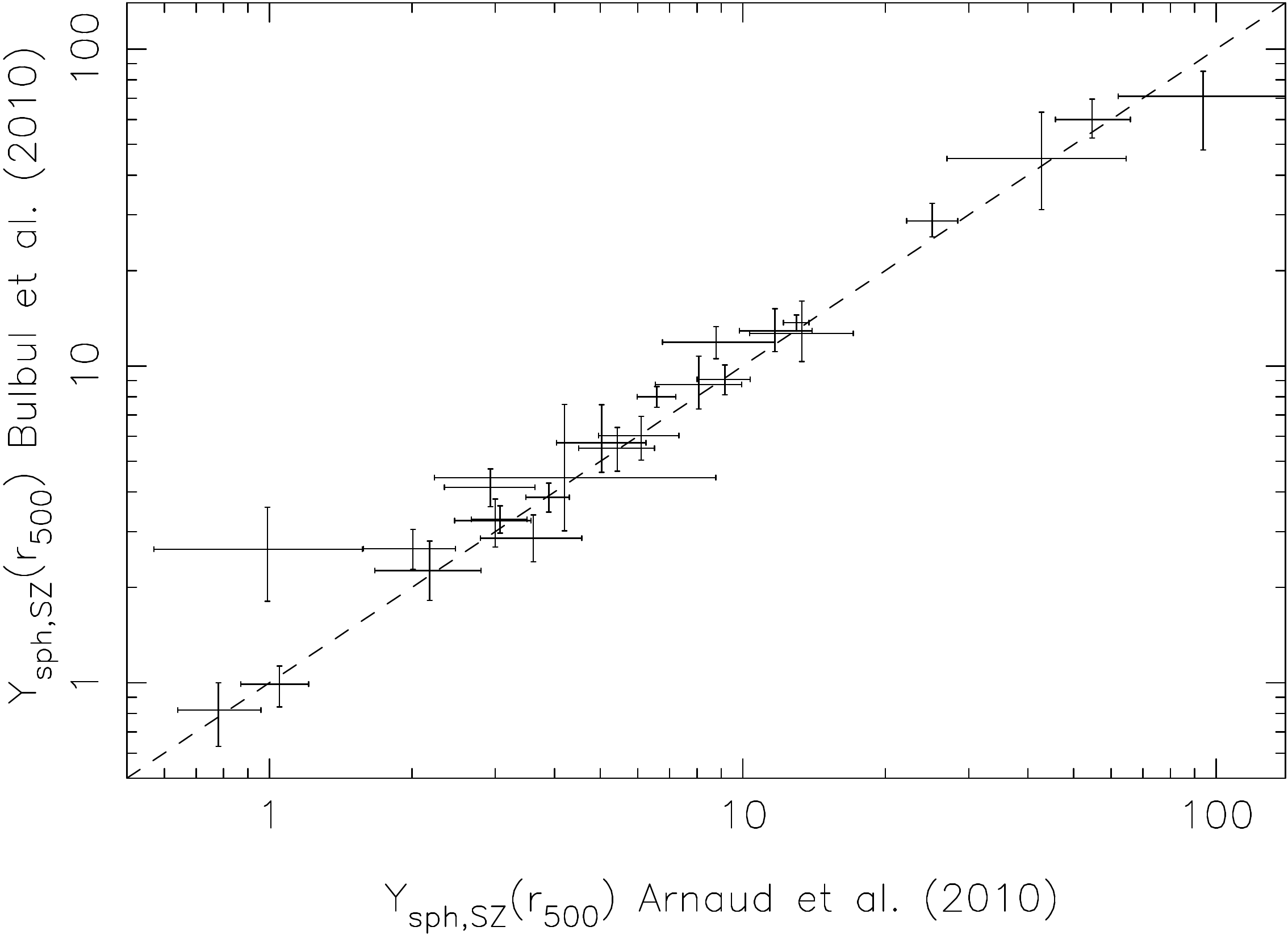}
\caption{Integrated SZ pressure $Y_{\rm sph, SZ}$ 
calculated using the \cite{bulbul2010} model ($y$ axis) and the \cite{arnaud2010} model ($x$ axis),
from a fit to the SZ data.
The value of $r_{500}$ was determined by the joint modelling of the SZ and X-ray observations, and it
is the same for both measurements.
The dashed line is the curve $y=x$.}
\label{fig:b10-a10-y}
\end{figure}

\begin{figure}[!t]
\centering
\includegraphics[width=3.5in,angle=-90]{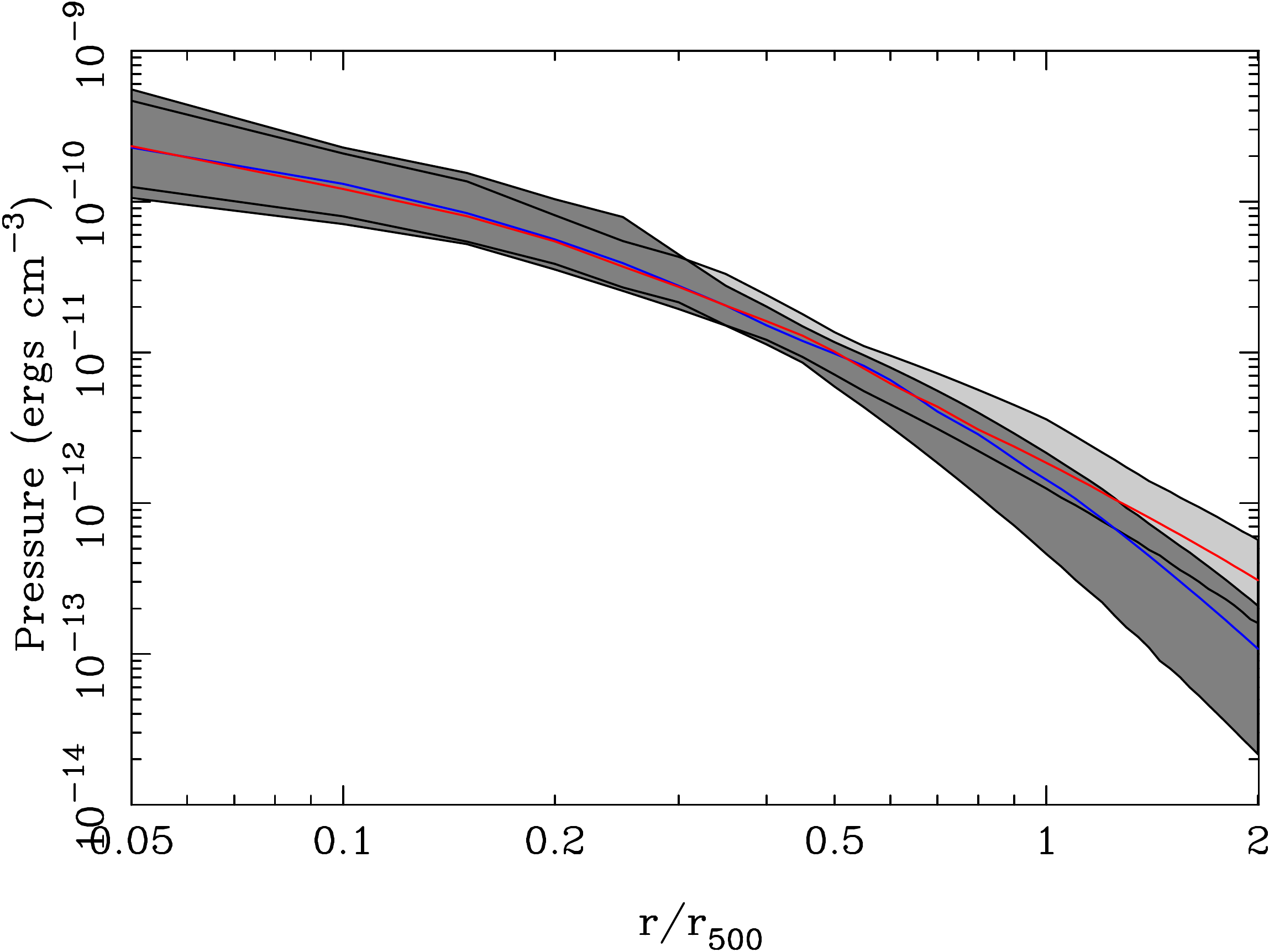}
\caption{Average pressure profiles from SZ fits to \cite{bulbul2010} model
(dark grey area, blue line), and to the \cite{arnaud2010} model (light grey area,
red line).
The lines are the
median of the 25 best-fit distributions, and the error bands
are the 68\% confidence level.}
\label{fig:b10-a10-average-profile}
\end{figure}

The \sza\ data were also fit to the \cite{arnaud2010} model using the same value
of $r_{500}$ as above. The best-fit parameters are shown in 
Table~\ref{tab:pres-params-a10}.  We compare the results from the 
\cite{bulbul2010} average pressure model with the \cite{arnaud2010} model in Figure~\ref{fig:b10-a10-y},
and find very good agreement: the weighted average of the ratio between 
the \cite{bulbul2010} and the  \cite{arnaud2010} models is 1.05$\pm$0.06.
A fit of the two measurements to a $y=x$ model assuming the values are independent
results in a $\chi^2_{min}$=5.6 for 25 degrees of freedom, consistent with 
the presence of negligible scatter between the two measurements.
The low value of $\chi^2_{min}$ is likely due to correlated errors, since the two measurements 
make use of the same data.
Figure~\ref{fig:b10-a10-average-profile} shows the average \citet{arnaud2010} and \citet{bulbul2010} 
pressure profiles for our sample.  The two parameterizations result in fits that are consistent
at all radii within $r_{500}$.
The consistency between the pressure profiles and the integrated $Y(r_{500})$ values
measured from the two models indicate that the choice of parameterization for the gas pressure
does not introduce a significant bias in the calculation of the integrated pressure within $r_{500}$.

\begin{table}[!t]
\centering
\scriptsize
\caption{Best-fit parameters for the fit of the \sza\ data to the  \cite{bulbul2010} average pressure model, and integrated
$Y$ parameter out to $r_{500}$. \label{tab:pres-params-b10}}
\begin{tabular}{lccc|c}
\hline
Cluster 	 & $P_{eo}$		   &    	$R_{s}$		  & 	$Y_{\rm sph,SZ,B10}$  &ratio ${Y_{\rm sph,SZ,B10}}/{Y_{\rm sph}} $ \\ 
                 &    ($10^{-11}$ ergs cm$^{-3}$) & (arcsec) & ($10^{-11}$) \\[0.2pc]
\hline
MACSJ0159.8-0849 &    	27.50$^{+15.59}_{-8.64}$ & 	40.99$^{+11.10}_{-9.58}$ & 	9.07$^{+1.01}_{-0.96}$ & 0.94$\pm$0.23 \\ 
Abell~383        &    	69.55$^{+20.41}_{-18.33}$ & 	22.13$^{+3.71}_{-2.71}$ & 	4.14$^{+0.59}_{-0.54}$ & 0.99$\pm$0.29 \\ 
MACS0329.7-0212  &    	342.00$^{+164.70}_{-189.90}$ & 	8.58$^{+3.80}_{-1.42}$ & 	2.65$^{+0.40}_{-0.37}$ & 1.02$\pm$0.32 \\ 
Abell~478        &    	29.21$^{+9.36}_{-2.84}$  &	112.80$^{+16.68}_{-30.96}$ & 	70.98$^{+14.10}_{-22.91}$&1.17$\pm$0.39\\ 
MACSJ0429.6-0253 &    	76.52$^{+105.50}_{-49.91}$  & 	158.00$^{+11.74}_{-5.10}$ & 	2.86$^{+0.53}_{-0.45}$ & 0.87$\pm$0.29 \\ 
3C186            &     	44.60$^{+6.61}_{-3.33}$& 	13.28$^{+1.55}_{-1.65}$ & 	0.82$^{+0.18}_{-0.19}$ & 0.95$\pm$0.32 \\ 
MACSJ0744.9+3927 &    	45.46$^{+46.37}_{-17.59}$ & 	29.91$^{+13.04}_{-10.73}$ & 	5.50$^{+0.90}_{-0.84} $& 1.45$\pm$0.43 \\ 
MACSJ0947.2+7623 &    	31.44$^{+23.66}_{-12.82}$  & 	34.31$^{+13.30}_{-9.49}$ & 	6.03$^{+0.90}_{-0.98}$ & 1.01$\pm$0.31 \\ 
Zwicky 3146      &    	103.20$^{+39.66}_{-25.83}$&   	25.36$^{+4.43}_{-4.13}$ & 	11.89$^{+1.43}_{-1.34} $&1.13$\pm$0.27 \\ 
MACSJ1115.8+0129 &   	168.10$^{+80.39}_{-49.95}$ & 	16.69$^{+3.27}_{-2.90}$ & 	7.99$^{+0.62}_{-0.58} $& 1.28$\pm$0.30 \\ 
MS1137.5+6625    &    	24.56$^{+23.35}_{-9.57}$&  	21.20$^{+6.99}_{-6.27}$ & 	0.99$^{+0.14}_{-0.15}$ & 1.36$\pm$0.37 \\ 
Abell~1413       &    	12.53$^{+2.33}_{-1.45}$ & 	147.10$^{+44.37}_{-38.53}$ & 	45.11$^{+18.10}_{-13.95}$&1.95$\pm$0.81\\ 
CLJ1226.9+3332   &    	140.50$^{+78.93}_{-67.19}$ & 	129.00$^{+5.15}_{-2.61}$ & 	3.28$^{+0.34}_{-0.31}$  &1.07$\pm$0.29 \\ 
MACSJ1311.0-0311 &    	7.56$^{+7.93}_{-5.26}$ &        54.26$^{+90.72}_{-21.33}$ & 	2.64$^{+0.94}_{-0.94}$  &1.14$\pm$0.47 \\ 
RXJ1347.5-1145   &    	296.71$^{+98.58}_{-58.19}$ & 	16.14$^{+1.96}_{-2.20}$ & 	13.70$^{+0.78}_{-0.82} $ &0.63$\pm$0.13\\ 
Abell~1835       &     	53.08$^{+18.43}_{-15.05}$ & 	47.82$^{+11.38}_{-8.06}$ & 	28.70$^{+3.90}_{-3.16}$  &0.97$\pm$0.22\\ 
MACSJ1423.8+2404 &    	36.51$^{+67.61}_{-24.81}$ &  	24.80$^{+36.43}_{-11.04}$ & 	4.44$^{+3.12}_{-1.42}$  & 1.76$\pm$1.03\\ 
MACSJ1427.3+4408 &    	90.97$^{+53.49}_{-54.50}$ &   	13.51$^{+8.13}_{-3.06}$ & 	2.26$^{+0.54}_{-0.44}$  & 0.48$\pm$0.14\\ 
RXJ1504.1-0248   &    	159.80$^{+141.20}_{-67.18}$ &  	22.95$^{+7.81}_{-6.35}$ & 	12.91$^{+2.25}_{-1.82}$ & 0.72$\pm$0.18\\ 
MACSJ1532.9+302  &    	44.70$^{+69.13}_{-27.53}$ & 	28.87$^{+28.77}_{-12.09}$ & 	5.72$^{+1.83}_{-1.10}$  & 1.24$\pm$0.42\\ 
MACSJ1621.6+3810 &    	14.56$^{+13.40}_{-4.74}$  &	41.13$^{+14.72}_{-14.84}$ & 	3.25$^{+0.55}_{-0.57}$  & 1.18$\pm$0.35\\ 
Abell~2204       &    	34.12$^{+6.11}_{-5.84}$  &	89.47$^{+16.45}_{-11.79}$ & 	59.98$^{+9.65}_{-7.56}$&  1.37$\pm$0.33\\ 
MACSJ1720.3+3536 &    	78.69$^{+50.98}_{-29.90}$ &  	18.54$^{+5.49}_{-4.26}$ & 	3.85$^{+0.42}_{-0.39} $ & 0.98$\pm$0.24\\ 
RXJ2129.6+0005   &    	21.23$^{+14.94}_{-8.42}$  &	56.85$^{+28.27}_{-16.83}$ & 	12.67$^{+3.37}_{-2.34}$ & 1.21$\pm$0.38\\ 
Abell~2537       &    	15.75$^{+8.35}_{-5.48}$ & 	55.18$^{+22.51}_{-14.36}$ & 	8.74$^{+1.99}_{-1.43}$  & 1.19$\pm$0.35\\ 
\hline

\end{tabular}
\end{table}

\begin{table}[!t]
\centering
\scriptsize
\caption{Best-fit parameters for the fit of the \sza\ data to the \cite{arnaud2010} model, and integrated
$Y$ parameter out to $r_{500}$. \label{tab:pres-params-a10}}
\begin{tabular}{lccc|c}
\hline
Cluster 	 & $p_{e,i}$		   &    	$r_{p}$		  & 		$Y_{\rm sph,SZ,A10}$  & ratio $Y_{\rm sph,SZ,B10}/Y_{\rm sph,SZ,A10}$\\ 
                 &    ($10^{-11}$ ergs cm$^{-3}$)& 	($^{\prime\prime}$)   & 	($10^{-11}$)  & \\
\hline
\\[-0.6pc]
MACSJ0159.8-0849 &    	6.38$^{+2.71}_{-1.88}$ & 	221.0$^{+56.9}_{-42.1}$ & 	9.16$^{+1.21}_{-1.15}$ & 0.99$\pm$0.17 \\[0.2pc] 
Abell~383        &    	48.1$^{+101~}_{-29.6}$ & 	~74.0$^{+37.7}_{-27.5}$ & 	2.93$^{+0.71}_{-0.59}$ &    1.41$\pm$0.37\\[0.2pc] 
MACS0329.7-0212  &    	~440$^{+787~}_{-328~}$ & 	~25.0$^{+17.7}_{-8.50}$ & 	2.01$^{+0.46}_{-0.43}$ &    1.32$\pm$0.35\\[0.2pc] 
Abell~478        &    	7.50$^{+1.43}_{-1.02}$ &	662.5$^{+209~}_{-153~}$ & 	93.9$^{+48.3}_{-31.8}$ &     0.76$\pm$0.42\\[0.2pc] 
MACSJ0429.6-0253 &    	3.50$^{+5.01}_{-1.95}$ & 	206.7$^{+161~}_{-81.9}$ & 	3.61$^{+0.96}_{-0.82}$ &    0.79$\pm$0.24\\[0.2pc] 
3C186            &     	1570$^{+609~}_{-706~}$ & 	~10.3$^{+3.11}_{-1.26}$ & 	0.78$^{+0.18}_{-0.14}$ &     1.05$\pm$0.32\\[0.2pc] 
MACSJ0744.9+3927 &    	11.5$^{+8.44}_{-4.34}$ & 	149.8$^{+63.7}_{-44.2}$ & 	5.43$^{+1.08}_{-0.93}$ &    1.01$\pm$0.25\\[0.2pc] 
MACSJ0947.2+7623 &    	6.68$^{+5.66}_{-2.79}$ & 	195.0$^{+81.4}_{-57.1}$ & 	6.10$^{+1.23}_{-1.14}$ &    0.99$\pm$0.25\\[0.2pc] 
Zwicky~3146      &    	39.4$^{+50.0}_{-21.0}$ &   	105.7$^{+51.3}_{-33.9}$ & 	8.78$^{+2.91}_{-2.02}$ &    1.35$\pm$0.41 \\[0.2pc] 
MACSJ1115.8+0129 &   	27.0$^{+12.3}_{-7.88}$ & 	108.7$^{+20.2}_{-17.8}$ & 	6.58$^{+0.64}_{-0.60}$ &    1.21$\pm$0.15\\[0.2pc] 
MS1137.5+6625    &    	20.3$^{+79.4}_{-14.5}$ &  	~61.3$^{+52.0}_{-30.2}$ & 	1.05$^{+0.16}_{-0.18}$ &    0.94$\pm$0.21\\[0.2pc] 
Abell~1413       &    	4.11$^{+0.77}_{-0.54}$ & 	619.0$^{+231~}_{-167~}$ & 	42.8$^{+21.8}_{-15.8}$ &     1.06$\pm$0.60\\[0.2pc] 
CLJ1226.9+3332   &    	35.9$^{+43.5}_{-19.0}$ & 	~67.6$^{+30.2}_{-20.3}$ & 	3.07$^{+0.43}_{-0.40}$ &    1.07$\pm$0.18\\[0.2pc] 
MACSJ1311.0-0311 &    	~709$^{+2240}_{-641~}$ & 	~16.2$^{+24.7}_{-7.56}$ & 	0.99$^{+0.58}_{-0.42}$ &    2.67$\pm$1.62\\[0.2pc] 
RXJ1347.5-1145   &    	50.7$^{+9.85}_{-7.97}$ & 	102.7$^{+9.48}_{-8.91}$ & 	13.0$^{+0.82}_{-0.80}$ &     1.06$\pm$0.09\\[0.2pc] 
Abell~1835       &     	14.7$^{+3.77}_{-3.11}$ & 	230.6$^{+36.3}_{-29.8}$ & 	25.1$^{+3.34}_{-2.94}$ &     1.14$\pm$0.20\\[0.2pc] 
MACSJ1423.8+2404 &    	7.82$^{+24.9}_{-5.44}$ &  	137.8$^{+239~}_{-75.2}$ & 	4.20$^{+4.57}_{-1.97}$ &    1.06$\pm$0.98\\[0.2pc] 
MACSJ1427.3+4408 &    	15.9$^{+32.4}_{-10.1}$ &   	~84.6$^{+58.5}_{-34.3}$ & 	2.18$^{+0.62}_{-0.51}$ &    1.04$\pm$0.35\\[0.2pc] 
RXJ1504.1-0248   &    	30.7$^{+18.9}_{-11.0}$ &   	136.6$^{+36.1}_{-28.9}$ & 	11.7$^{+2.31}_{-1.85}$ &    1.10$\pm$0.29\\[0.2pc] 
MACSJ1532.9+302  &    	12.8$^{+9.26}_{-4.87}$ & 	135.8$^{+48.2}_{-36.3}$ & 	5.03$^{+1.21}_{-0.99}$ &    1.14$\pm$0.38\\[0.2pc] 
MACSJ1621.6+3810 &    	4.78$^{+2.74}_{-1.59}$ &	171.9$^{+61.1}_{-43.5}$ & 	3.00$^{+0.57}_{-0.54}$ &    1.08$\pm$0.27\\[0.2pc] 
Abell~2204       &    	9.88$^{+1.85}_{-1.51}$ &	416.4$^{+72.8}_{-59.0}$ & 	54.7$^{+11.2}_{-8.98}$ &    1.10$\pm$0.26\\[0.2pc] 
MACSJ1720.3+3536 &    	13.1$^{+5.22}_{-3.52}$ &  	119.5$^{+22.2}_{-19.4}$ & 	3.89$^{+0.41}_{-0.41}$ &     0.99$\pm$0.15\\[0.2pc] 
RXJ2129.6+0005   &    	4.50$^{+2.45}_{-1.39}$ &	328.6$^{+128~}_{-89.2}$ & 	13.3$^{+3.80}_{-2.98}$ &    0.95$\pm$0.32\\[0.2pc] 
Abell~2537       &    	4.63$^{+2.29}_{-1.37}$ & 	250.6$^{+80.5}_{-59.4}$ & 	8.07$^{+1.87}_{-1.54}$ &    1.08$\pm$0.31\\[0.2pc] 
\hline

\end{tabular}
\end{table}

\section{Discussion}
\label{sec:discussion}


The agreement we find between SZ and X-ray measurements of the $Y_{\rm sph}(r_{500})$ parameter
is consistent with a simple scenario in which  the SZ decrement  and the X-ray emission  
from massive relaxed clusters
originate from the same highly-ionized thermal plasma, with only small
contributions from other possible sources of emission.
This result is in agreement with earlier $\sim 30$~GHz SZ studies using the 
Owens Valley Radio Observatory (OVRO) and the Berkeley Illinois Maryland Array (BIMA) millimeter arrays, in which the same
value of the gas mass fraction was measured using SZ and X-ray data \citep{laroque2006}
Our results also support the findings by
\cite{melin2011} and \cite{Planck2011-sz} of an overall agreement between
the two measurements of the thermal pressure.

We find scatter between the \sza\ and \chandra\ $Y_{\rm sph}$ estimates at
a level of 16\%.  A possible source 
of systematic error that could give rise to this scatter, and that is particularly relevant to our 
measurements out to $r_{500}$, is elongation of the cluster along the line of sight.
We use spherically symmetric models in the analysis; an intrinsically prolate cluster 
(elongated along the line of sight), when fit to a spherical model,
will have its X-ray surface brightness --- and therefore the corresponding
$Y_{\rm sph}$ parameter --- underestimated with respect to the corresponding SZ measurement 
\citep[e.g.,][]{cooray2000,defilippis2005,ameglio2007}. This is due to the quadratic
dependence of the X-ray surface brightness profile on the density, as opposed
to the linear dependence of the SZ effect.
Our sample has just three clusters with a statistically significant deviation 
from the $Y_{\rm sph}=Y_{\rm sph,SZ}$ line, but
in the direction of $Y_{\rm sph}/Y_{\rm sph,SZ}>1$, and therefore consistent with oblateness
(compression along the line of sight) rather than prolateness. 
The fact that the \cite{allen2008} sample of relaxed clusters is X-ray selected may lead to 
including preferentially oblate clusters as their surface brightness will be 
boosted.  An alternative interpretation for the presence
of scatter between the \sza\ and \chandra\ estimates of $Y$ is that some of these clusters are
disturbed and have undergone a recent merger, as is
almost certainly the case for RXJ1347.5-1145 \citep{mason2010,johnson2011}.
A merger would result in clumping of the gas, and therefore an overstimate
of the gas mass and $Y$ from X-ray measurements, as suggested
by \cite{simionescu2011} to explain the observations of the Persues
cluster. Clumping would not affect the SZ observations, because
of the linear dependence of the signal on density.


The  fit of the SZ data to the universal
pressure profile of \cite{arnaud2010}, and to the
 average pressure
profile based on the \cite{bulbul2010} parameterization of the pressure,
are statistically acceptable for all clusters, with a similar $\chi^2$ for the two
models.
The agreement between $Y_{\rm sph}$ at $r_{500}$ using the 
two models indicates that
the integrated pressure is not highly sensitive to (reasonable)
choices of parameterization.  

We have adopted throughout our analysis the value of $r_{500}$ determined
from the joint SZ and X-ray observations. In the absence of X-ray information,
one may instead adopt a fiducial value of the gas mass fraction $f_{gas}$ to determine
$r_{500}$ \citep[e.g.,][]{joy2001,bonamente2008,mroczkowski2011}, or other means based on SZ--mass 
scaling relations.  The additional assumptions required to estimate $r_{500}$ from SZ data only 
will likely contribute additional scatter to the $Y_{\rm sph} - Y_{\rm sph,SZ}$ relation, 
when $r_{500}$ used to measure $Y_{\rm sph,SZ}$ is estimated directly from the SZ data.


\section{Conclusions}
\label{sec:conclusions}
We have presented the joint analysis of \sza\ and \chandra\ observations 
of the \cite{allen2008} sample of massive and relaxed galaxy clusters.
We have collected sensitive SZ data for all clusters at declination $\geq -15^\circ$
with no significant contamination from foreground or intrinsic radio sources, for a 
total of 25 clusters in the redshift range $0.09 \leq z \leq 1.06$.
We also used the X-ray imaging and spectroscopic \chandra\ data 
that are available for all clusters, and performed a cluster-by-cluster
comparison of the integrated pressure.  The $Y_{\rm sph}$ value estimated
from the joint SZ and X-ray data, and from the SZ data alone,
agree within a few percent at $r_{500}$, indicating that the SZ and X-ray signal from
massive relaxed clusters is consistent with a common thermal origin.
We therefore confirm the findings of \cite{melin2011} and \cite{Planck2011-sz}, and
find no evidence for the presence of significant sources of
systematic uncertainty in the measurements of the ICM pressure from 
SZ and X-ray observations of massive relaxed clusters.

We also determined an average pressure profile based on
the \cite{bulbul2010} model, with shape parameters ($n=3.5$ and $\beta=2.0$) determined
by a joint fit to \chandra\ X-ray data and our \sza\ observations of the \cite{allen2008}
sample of massive relaxed clusters. We have shown that measurements of
the radial profile of the pressure out to $r_{500}$, and of  $Y_{\rm sph, SZ}$ at 
$r_{500}$, agree between the \cite{arnaud2010} and the \cite{bulbul2010} average pressure profiles
out to $r_{500}$.  Our conclusions indicate that both models are adequate for describing cluster 
radial pressure profiles and measuring the integrated thermal energy content 
in relaxed clusters.

\section{Acknowledgments}
The operation of the SZA is supported by NSF through grant AST-0604982 and AST-0838187.  
Partial support is also provided from grant PHY-0114422 at the University of Chicago, 
and by NSF grants AST-0507545 and AST-05-07161 to Columbia University.  
CARMA operations are supported by the NSF under a cooperative
agreement, and by the CARMA partner universities.
Support for TM was provided by NASA through Einstein
Postdoctoral Fellowship grant number PF0-110077 awarded by the
Chandra X-ray Center, which is operated by the Smithsonian
Astrophysical Observatory for NASA under contract NAS8-03060.


\bibliographystyle{apj}

\end{document}